\documentclass[iop]{emulateapj}
\usepackage{epsfig,amsfonts,amsmath,graphicx,natbib,apjfonts}
\pdfoutput=1
\def\ra#1#2#3{#1$^{\rm h}$#2$^{\rm m}$#3$^{\rm s}$}
\def\dec#1#2#3{$#1^\circ#2'#3''$}
\def\nod{\nodata}
\def\hst{{\it HST}}
\def\swift{{\it Swift}}

\def\har{1}

\shorttitle{Short GRB locations}
\shortauthors{Fong \& Berger}

\begin{document}

\title{The Locations of Short Gamma-ray Bursts as Evidence for Compact Object Binary Progenitors}

\author{ 
W.~Fong\altaffilmark{\har} \&
E.~Berger\altaffilmark{\har}
}

\altaffiltext{1}{Harvard-Smithsonian Center for Astrophysics, 60
Garden Street, Cambridge, MA 02138, USA}

\begin{abstract}

We present a detailed investigation of {\it Hubble Space Telescope}
rest-frame UV/optical observations of 22 short gamma-ray burst (GRB)
host galaxies and sub-galactic environments. Utilizing the high
angular resolution and depth of \hst\, we characterize the host galaxy
morphologies, measure precise projected physical and host-normalized
offsets between the bursts and host centers, and calculate the
locations of the bursts with respect to their host light distributions
(rest-frame UV and optical). We calculate a median short GRB projected
physical offset of $4.5$~kpc, about $3.5$ times larger than that for
long GRBs, and find that $\approx 25\%$ of short GRBs have offsets of
$\gtrsim 10$~kpc. When compared to their host sizes, the median offset
is $1.5$~half-light radii ($r_e$), about $1.5$ times larger than the values
for long GRBs, core-collapse supernovae, and Type Ia supernovae. In addition,
$\approx 20\%$ of short GRBs having offsets of $\gtrsim 5r_e$, and
only $\approx 25\%$ are located within $1r_e$. We further find that short GRBs
severely under-represent their hosts' rest-frame optical and UV light,
with $\approx 30-45\%$ of the bursts located in regions of their host
galaxies that have no detectable stellar light, and $\approx 55\%$ in
the regions with no UV light. Therefore, short GRBs do not occur in
regions of star formation or even stellar mass. This demonstrates that
the progenitor systems of short GRBs must migrate from their birth
sites to their eventual explosion sites, a signature of kicks in
compact object binary systems. Utilizing the full sample of offsets,
we estimate natal kick velocities of $\approx
20-140$~km~s$^{-1}$. These independent lines of evidence provide the
strongest support to date that short GRBs result from the merger of
compact object binaries (NS-NS/NS-BH).

\end{abstract}


\section{Introduction}

The environments of cosmic explosions and their locations within their hosts provide invaluable insight on
the nature of their progenitors. For instance, the spatial locations
of long gamma-ray bursts (GRBs; $T_{90} \gtrsim 2$~s;
\citealt{kmf+93}) within their exclusively star-forming host galaxies are consistent with the
expected distribution for massive stars in exponential disks
\citep{bkd02}. The result is similar for core-collapse supernovae
(SNe; \citealt{psb08}), which are only found in spiral and irregular
galaxies \citep{vlf05,hpm+08,lcl+11}, indicative of a young, massive
star origin. In contrast, Type Ia SNe originate in both star-forming
and elliptical galaxies \citep{ot79,vlf05,mdp+05,lcl+11}, and their locations do not coincide with star-forming regions
\citep{psb08,wwf+13}, consistent with an evolved progenitor system.

Equally important to the spatial offsets are the locations of these
transients with respect to the distribution of the underlying host galaxy
light. Using {\it Hubble Space Telescope (HST)}
observations, it has been shown that long GRBs are concentrated on the
brightest ultra-violet (UV) regions of their host galaxies
\citep{fls+06,slt+10}, pointing to explosion sites within
unusually bright star-forming regions. Similarly, core-collapse SNe
tend to explode in bright UV regions within their hosts
\citep{kkp08,slt+10}. In contrast, Type Ia SNe under-represent
their hosts' rest-frame UV light, suggesting that they do not tend to
occur in regions of star formation \citep{kkp08,wwf+13}, but may
correlate with optical light, suggesting a dependence on
stellar mass \citep{kkp08,wwf+13} as expected for their white dwarf
progenitors.

For short GRBs ($T_{90} \lesssim 2$~s), several theoretical
progenitor systems have been proposed, including NS-NS/NS-BH mergers
\citep{elp+89,npp92}, accretion-induced-collapse of a WD or NS
\citep{qwc+98,lwc+06,mqt08} and magnetar flares
\citep{lwc+06,clw+08}. Studies of the short GRB host galaxy
demographics have shown that $\approx 1/4$ of these events explode in
elliptical galaxies with no signs of star formation
\citep{fbc+13}, but the majority occur in star-forming galaxies \citep{ber09,fbc+13}. The inferred
progenitor ages are $\approx 0.1-$few~Gyr \citep{lb10,fbc+13},
pointing to an origin from older stellar populations.

In \citealt{fbf10}, we used \hst\ observations of ten short GRB host
galaxies to study their sub-galactic environments. For the seven bursts with sub-arcsecond localizations, and
thus robust associations to a host galaxy, we constrained their host
morphologies, spatial and host-normalized offsets, light
distributions, and compared these results to the distributions for
long GRBs. We found spatial offsets that are five times greater than
those for long GRBs, but with a similar median host-normalized offset
of $\approx 1$~half-light radius. In addition, we found preliminary
evidence that short GRBs under-represent their hosts' rest-frame
optical and UV light, in stark contrast to long GRBs and core-collapse
SNe. Due to the small number of events, the light distribution results
were only suggestive, and not statistically significant. Separately,
we also used \hst\ data to study an emerging population of bursts with
no obvious coincident host galaxy to optical limits of $\gtrsim
26$~mag and found that these bursts likely have large offsets of
$30-100$~kpc from their hosts \citep{ber10}.

These initial studies demonstrate that \hst\ observations are
essential for characterizing the local environments of short GRBs, and
thus their progenitors. The angular resolution is important in
measuring precise offsets and locating the afterglow to sub-pixel
precision, critical to analyzing the placement within the host light
distributions. In addition, the depth of \hst\ allows for the
potential detections of faint coincident host galaxies, surpassing the
capabilities of ground-based instuments in the optical and and
near-infrared.

Expanding on this initial work, we present here \hst\ observations of
$16$ additional short GRB host galaxies, $15$ of which have
sub-arcsecond positions. We describe the data reduction procedures,
including photometry, astrometry, surface brightness profile fits,
offsets and fractional flux determination in Section~\ref{sec:dr}. We
combine the results with those from \citet{fbf10} and analyze the
entire sample of $22$ events in Section~\ref{sec:analysis}. Finally, in
Section~\ref{sec:disc}, we
consider the implications for the progenitor systems 

Throughout the paper we use the standard cosmological parameters,
$H_0=71$ km s$^{-1}$ Mpc$^{-1}$, $\Omega_m=0.27$, and
$\Omega_\Lambda=0.73$.  All reported magnitudes are corrected for
Galactic extinction using dust maps \citep{sf11} and are calibrated to
the AB magnitude system.

\tabletypesize{\scriptsize}
\begin{deluxetable*}{lcccccccccc}
\tablecolumns{11}
\tabcolsep0.0in\footnotesize
\tablewidth{0pc}
\tablecaption{{\it HST} Observations of Short GRB Host Galaxies
\label{tab:hst_obs}}
\tablehead{
\colhead{GRB}                  &
\colhead{RA}                   &
\colhead{Dec}                  &
\colhead{Uncert.}              &
\colhead{$z$}                  &
\colhead{Instrument}           &
\colhead{Filter}               &
\colhead{Date}                 &
\colhead{Exp.~Time}            &
\colhead{AB mag$^{a}$}         &
\colhead{$A_{\lambda}$}    \\           
\colhead{}                     &
\colhead{(J2000)}              &
\colhead{(J2000)}              &
\colhead{($''$)}               &
\colhead{}                     &
\colhead{}                     &
\colhead{}                     &
\colhead{(UT)}                 &
\colhead{(s)}                  &
\colhead{}                     &
\colhead{(mag)}                           
}
\startdata
061201  & \ra{22}{08}{32.13} & \dec{-74}{34}{47.05} & $0.12$ & $0.111?$ & WFC3/IR   & F160W & 2012 Jun 11 & 6397 & $>26.4 / 18.63 \pm 0.01$ / $24.46 \pm 0.03^{b}$ & $0.039$ \\
070429B & \ra{21}{52}{03.81} & \dec{-38}{49}{42.0}  & $1.5^{c}$ & $0.9023$ & WFC3/IR & F160W & 2010 Apr 26 & 2797 & $20.59 \pm 0.03$ & $0.012$ \\
        &                    &                      &        &          & WFC3/UVIS & F475W & 2010 Apr 26 & 2797 & $24.31 \pm 0.20$ & $0.096$ \\
070707  & \ra{17}{50}{58.59} & \dec{-68}{55}{27.59} & $0.16$ & $<3.6$   & WFC3/IR   & F160W & 2010 Jun 24 & 6397 & $26.16 \pm 0.24$ & $0.035$ \\
        &                    &                      &        &          & ACS       & F606W & 2010 Jun 30 & 5644 & $26.68 \pm 0.12$ & $0.177$ \\
070714B & \ra{03}{51}{22.28} & \dec{+28}{17}{51.75} & $0.20$ & $0.923$  & WFC3/IR   & F160W & 2009 Aug 16 & 2798 & $22.99 \pm 0.02$ & $0.069$ \\
        &                    &                      &        &          & WFC3/UVIS & F475W & 2009 Aug 16 & 2698 & $24.89 \pm 0.06$ & $0.467$ \\
070724A & \ra{01}{51}{14.10} & \dec{-18}{35}{39.28} & $0.08$ & $0.4571$ & WFC3/IR   & F160W & 2011 Oct 3  & 2396 & $19.89 \pm 0.02$ & $0.006$ \\
070809  & \ra{13}{35}{04.55} & \dec{-22}{08}{30.8}  & $0.40$  & $0.473?$ & WFC3/IR  & F160W & 2010 May 5  & 5597 & $>26.2 / 18.22 \pm 0.01^{d}$ & $0.041$ \\
        &                    &                      &        &          & ACS       & F606W & 2009 Aug 9  & 5150 & $>28.1 / 20.47 \pm 0.03^{d}$ & $0.210$ \\
071227  & \ra{03}{52}{31.25} & \dec{-55}{59}{02.63} & $0.22$ & $0.381$  & WC3/IR    & F160W & 2010 Jun 11 & 2797 & $18.73 \pm 0.01$ & $0.006$ \\
        &                    &                      &        &          & WFC3/UVIS & F438W & 2010 Jun 12 & 2900 & $22.35 \pm 0.05$ & $0.042$ \\
080503  & \ra{19}{06}{28.77} & \dec{+68}{47}{35.32} & $0.14$ & $<4.2$   & WFC3/IR   & F160W & 2011 Dec 26 & 5597 & $>26.2 / 25.84 \pm 0.07^{d}$ & $0.028$ \\
080905A & \ra{19}{10}{41.74} & \dec{-18}{52}{47.44} & $0.18$ & $0.1218$ & WFC3/IR   & F160W & 2011 Oct 16 & 2397 \\
        &                    &                      &        &          & WFC3/IR   & F160W & 2012 Apr 14 & 2397 & $25.97 \pm 0.11^{ef}$ & $0.014$ \\
        &                    &                      &        &          & WFC3/UVIS & F814W & 2012 Apr 14 & 2600 & $>27.5^{f}$          & $0.040$ \\
        &                    &                      &        &          & WFC3/UVIS & F606W & 2012 Apr 14 & 2600 & $27.29 \pm 0.14^{f}$ & $0.066$ \\
090305A & \ra{16}{07}{07.59} & \dec{-31}{33}{22.12} & $0.20$ & $<4.1$   & WFC3/IR   & F160W & 2012 Feb 22 & 5597 & $25.20 \pm 0.10$ & $0.092$ \\
090426  & \ra{12}{36}{18.05} & \dec{+32}{59}{09.42} & $0.12$ & $2.609$  & WFC3/IR   & F160W & 2011 Oct 28 & 2397 & $25.56 \pm 0.07$ & $0.008$ \\
090510  & \ra{22}{14}{12.53} & \dec{-26}{34}{59.0}  & $0.20$  & $0.903$   & WFC3/IR  & F160W & 2011 Oct 11 & 2397 & $21.79 \pm 0.01$ & $0.009$ \\
090515  & \ra{10}{56}{36.10} & \dec{+14}{26}{29.37} & $0.16$ & $0.403?$ & WFC3/IR   & F160W & 2011 Oct 24 & 5597 & $>26.1 / 18.42 \pm 0.02^{d}$ & $0.001$ \\
091109B & \ra{07}{30}{56.61} & \dec{-54}{05}{23.11} & $0.16$ & $<4.4$   & WFC3/IR   & F160W & 2012 Feb 26 & 5596 & $>25.0 / 19.74 \pm 0.03^{d}$ & $0.075$ \\
100117A & \ra{00}{45}{04.65} & \dec{-01}{35}{41.99} & $0.16$ & $0.915$  & WFC3/IR   & F160W & 2011 Sep 29 & 2397 & $21.37 \pm 0.04$ & $0.011$ \\
130603B & \ra{11}{28}{48.17} & \dec{+17}{04}{18.03} & $0.09$ & $0.3564$ & WFC3/IR   & F160W & 2013 Jun 13 & 2612 & $19.83 \pm 0.02$ & $0.012$ \\
        &                    &                      &        &          & ACS       & F606W & 2013 Jun 13 & 2216 & $21.08 \pm 0.04$ & $0.057$
\enddata
\tablecomments{$^{a}$ Corrected for Galactic extinction, $A_{\lambda}$ \citep{sf11}. \\
$^{b}$ For GRB\,061201, we report the $3\sigma$ limit on a coincident host galaxy, photometry for ``G1'' and for ``G2'', respectively. \\
$^{c}$ $1\sigma$ positional error radius from \swift/XRT \citep{gtb+07,ebp+09} \\ 
$^{d}$ For GRBs\,070809, 080503, 090515 and 091109B, we report both the $3\sigma$ limit on a coincident host galaxy and the magnitude of the galaxy with the lowest probability of chance coincidence. \\
$^{e}$ To attain a higher signal-to-noise ratio, photometry is reported for the 2011 October 16 and 2012 April 14 observations combined. \\
$^{f}$ The position of ``G1'' is contaminated with saturated stars, so photometry is reported here only for ``G2''. The F814W limit corresponds to $3\sigma$.}
\end{deluxetable*}

\section{Data Reduction}
\label{sec:dr}

\subsection{Sample}

We study {\it HST} observations of $16$ short GRB host galaxies and
environments obtained with the Advanced Camera for Surveys (ACS/WFC)
and the infrared and ultraviolet-visual channels on the Wide-Field
Camera~3 (WFC3/IR and WFC3/UVIS). The data were obtained as part of
programs 11669 and 12502 (PI:~Fruchter), which targeted all short GRBs
with optical positions from April 2007 to January 2010, and the
Director's Discretionary Time program 13497 (PI:~Tanvir) for
GRB\,130603B. We combine these public \hst\ data with ground-based
observations of optical afterglows to astrometrically locate the burst
positions within the host galaxies.

Ten of the 16 bursts have apparent host galaxies based on ground-based
imaging (GRBs\,070429B, 070707, 070714B, 070724A, 071227, 080905A,
090426A, 090510, 100117A, and 130603B), and all of these except
GRB\,070707 have spectroscopic redshifts
\citep{cbn+08,pdc+08,bcf+09,dmc+09,gfl+09,ktr+10,lbb+10,mkr+10,rwl+10,fbc+11,cpp+13}. We
note that for GRB\,080905A, the host association is less secure
(probability of chance coincidence, $P_{cc}(<\delta R) \approx 0.01$)
than the remaining bursts with $P_{cc}(<\delta R) \approx
10^{-4}-10^{-3}$ (\citealt{fbc+13} and Section~\ref{sec:pcc}), due to
the large separation from its claimed host galaxy \citep{rwl+10}. All
of these hosts (except GRB\,070429B) have reported near-infrared (NIR)
detections or limits
\citep{pdc+08,bcf+09,gfl+09,lb10,lbb+10,rwl+10,fbc+11,bfc13,tlf+13}.

The remaining six short GRBs have no known coincident host galaxies to
optical limits of $\gtrsim 26$~mag from previous ground-based or previous {\it
HST} observations (GRBs\,061201, 070809, 080503, 090305A, 090515, and
091109B; \citealt{pbm+08,pmg+09,ber10,fbf10,rot+10,gcn10154}). These
events have been termed ``host-less'', but appear to have host
galaxies at separations of $\approx 30-100$~kpc with low probability
of chance coincidence \citep{ber10}.

With the exception of GRB\,130603B \citep{bfc13,tlf+13}, all of the
\hst\ observations presented here have not been published in the
literature thus far. The results in this work, combined with the \hst\
data from \citet{fbf10}, GRB\,080503 \citep{pmg+09,ber10}, and
GRB\,130603B \citep{bfc13,tlf+13}, comprise the full available sample
of short GRB hosts with \hst\ observations. Details of the GRB
properties and the observations are provided in
Table~\ref{tab:hst_obs}.

\begin{figure*}[ht]
\begin{minipage}[c]{\textwidth}
\tabcolsep0.0in
\includegraphics*[width=0.25\textwidth,clip=]{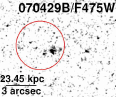} 
\includegraphics*[width=0.25\textwidth,clip=]{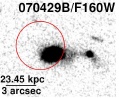} 
\includegraphics*[width=0.25\textwidth,clip=]{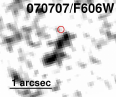} 
\includegraphics*[width=0.25\textwidth,clip=]{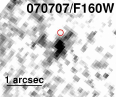} \\
\includegraphics*[width=0.25\textwidth,clip=]{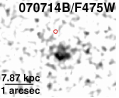} 
\includegraphics*[width=0.25\textwidth,clip=]{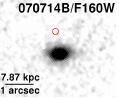} 
\includegraphics*[width=0.25\textwidth,clip=]{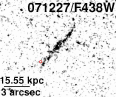} 
\includegraphics*[width=0.25\textwidth,clip=]{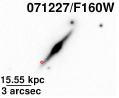} \\
\includegraphics*[width=0.25\textwidth,clip=]{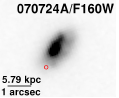} 
\includegraphics*[width=0.25\textwidth,clip=]{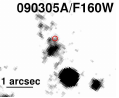} 
\includegraphics*[width=0.25\textwidth,clip=]{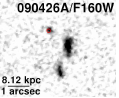} 
\includegraphics*[width=0.25\textwidth,clip=]{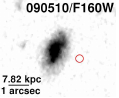} \\
\includegraphics*[width=0.25\textwidth,clip=]{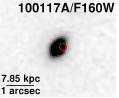} 
\includegraphics*[width=0.25\textwidth,clip=]{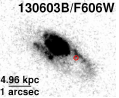} 
\includegraphics*[width=0.25\textwidth,clip=]{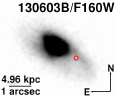}
\end{minipage}
\caption{{\it HST} observations of ten short GRBs with robust
associations to a host galaxy ($P_{cc}(<\delta R) \approx
10^{-4}-10^{-3}$) based on previous ground-based observations. The
afterglow positions are represented by a $3\sigma$ error circle in
each frame, except for GRBs\,070714B and 090510, where the afterglow
positional uncertainties are larger, and the circles correspond to
$1\sigma$. For GRB\,070429B, the position of the X-ray afterglow from
\swift/XRT is shown (red circle; $90\%$ containment,
\citealt{gtb+07,ebp+09}). For GRB\,130603B/F160W, the image after PSF
subtraction of the point source associated with the GRB (see
Section~\ref{sec:phot} and \citealt{bfc13}) is shown. All images are
oriented with North up and East to the left.
\label{fig:image}}
\end{figure*}

\begin{figure*}
\begin{minipage}[c]{\textwidth}
\tabcolsep0.0in
\includegraphics*[width=0.25\textwidth,clip=]{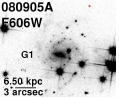} 
\includegraphics*[width=0.25\textwidth,clip=]{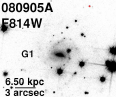} 
\includegraphics*[width=0.25\textwidth,clip=]{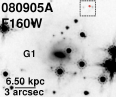} 
\includegraphics*[width=0.25\textwidth,clip=]{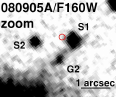}
\end{minipage}
\caption{{\it HST} observations of GRB\,080905A with the optical
afterglow position ($3\sigma$; red circle) indicated. The face-on
spiral galaxy at $z=0.1218$, claimed as the host by \citet{rwl+10}, is
labeled as ``G1'' ($P_{cc}(<\delta R) \approx 0.01$) in each of the 3
filters, while a zoomed version of the F160W observation shows a new
extended source, ``G2'' ($P_{cc}(<\delta R) \approx 0.08$), as well as
two sources with stellar PSFs denoted as ``S1'' and ``S2''. All images
are oriented with North up and East to the left.
\label{fig:080905}}
\end{figure*}

\begin{figure*}
\begin{minipage}[c]{\textwidth}
\includegraphics*[width=0.25\textwidth,clip=]{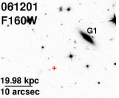} 
\includegraphics*[width=0.25\textwidth,clip=]{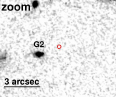} 
\includegraphics*[width=0.25\textwidth,clip=]{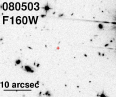} 
\includegraphics*[width=0.25\textwidth,clip=]{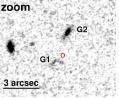} \\
\includegraphics*[width=0.25\textwidth,clip=]{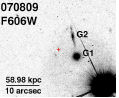} 
\includegraphics*[width=0.25\textwidth,clip=]{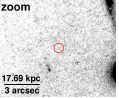} 
\includegraphics*[width=0.25\textwidth,clip=]{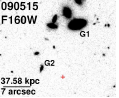} 
\includegraphics*[width=0.25\textwidth,clip=]{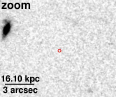} \\
\includegraphics*[width=0.25\textwidth,clip=]{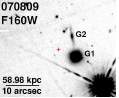} 
\includegraphics*[width=0.25\textwidth,clip=]{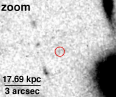} 
\includegraphics*[width=0.25\textwidth,clip=]{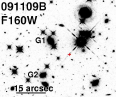} 
\includegraphics*[width=0.25\textwidth,clip=]{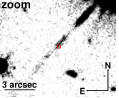}
\end{minipage}
\caption{{\it HST} observations of $5$ short GRBs with no coincident
host galaxy to $m_{\rm 160W} \gtrsim 26.2$ mag (``host-less''
bursts). We note that the afterglow position of GRB\,091109B is
contaminated by a diffraction spike so we place a comparatively
shallow limit on a coincident host galaxy of $m_{\rm 160W} \gtrsim 25$
mag. The large-scale environments (left) and the $10''$ surrounding
the afterglow position (right) are shown for each burst. The most
probable and second most probable host galaxies from probability of
chance coincidence analysis are labeled (``G1'' and ``G2'',
respectively). The afterglow positions are shown by the red cross or
error circle in each frame. Error circles are $5\sigma$ in radius
except for GRB\,070809, which is $1\sigma$ because the uncertainty is
based on absolute astrometry. Physical scales in kpc are based on the
redshift of ``G1'' for each burst, if known. All images are oriented
with North up and East to the left.
\label{fig:himage}}
\end{figure*}

\subsection{Image Processing}
\label{sec:datared}

We retrieved pre-processed images for
the 16 short GRBs from the \hst\
archive\footnotemark\footnotetext{http://archive.stsci.edu/hst/.}. We apply distortion corrections and combine the
individual exposures using the {\tt astrodrizzle} package in PyRAF
(Gonzaga~et.~al.~2012). For the ACS images we use {\tt
pixfrac}\,=\,1.0 and {\tt pixscale}\,=\,$0.05''$ pixel$^{-1}$. For the
WFC3/IR images, we use the recommended values of {\tt pixfrac}\,=\,1.0
and {\tt pixscale}\,=\,$0.0642''$ pixel$^{-1}$, half of the native
pixel scale, while for the WFC3/UVIS images, we use {\tt
pixscale}\,=\,$0.033''$ pixel$^{-1}$. The final drizzled images are shown for ten
events with established host galaxies (Figure~\ref{fig:image}),
multi-band observations of GRB\,080905A (Figure~\ref{fig:080905}) and
five events termed as ``host-less'' (Figure~\ref{fig:himage}).

We obtained public optical/NIR afterglow discovery images for each
burst. These images are from the UV-Optical Telescope (UVOT) on-board
the \swift\ satellite, the twin $6.5$-m Magellan telescopes, the
Gemini-North and South $8$-m telescopes, and the $8$-m Very Large
Telescope (VLT). The telescope and instrument for each afterglow image
is listed in Table~\ref{tab:offsets}. For the observations from
Magellan, Gemini, and VLT, we use standard IRAF tasks to process the
data. For GRBs\,070429B and 070714B, which have reported UVOT
afterglows \citep{gcn7145,gcn6689}, we use the {\tt uvotimsum} task as
part of the HEASOFT package to create co-added images for each of the
seven UVOT filters.

We confirm the afterglow detection of GRB\,070714B in multiple filters
\citep{gcn6689}.  For GRB\,070429B, the afterglow is reported to be
weak ($3.9\sigma$) and detectable only by combining the data from all
filters at $\delta t \approx 600-2660$~s. To assess whether this is a
real detection, we combine the UVOT observations in the same manner as
described in \citet{gcn7145}, but we do not detect any source at the
reported position, nor within the XRT error circle. We therefore do
not consider the reported afterglow to be real, and use the XRT
position (Table~\ref{tab:hst_obs}) in our analysis presented here.

\subsection{Photometry}
\label{sec:phot}

We perform aperture photometry for the galaxy with the lowest
probability of chance coincidence in each field (see
Section~\ref{sec:pcc}) using standard tasks in IRAF and the tabulated
zeropoints for ACS, WFC3/IR and WFC3/UVIS calibrated to the AB
magnitude system (Table~\ref{tab:hst_obs}). In addition, for the bursts
previously termed as ``host-less'' (GRBs\,061201, 070809, 080503,
090515, and 091109B; Figure~\ref{fig:himage}), we determine the
$3\sigma$ limit at the afterglow position.

We note that for GRB\,080905A, we can only perform photometry for
``G2'', because ``G1'', claimed to be the host galaxy by
\citet{rwl+10}, is contaminated by foreground saturated stars which we
cannot reliably subtract (Figure~\ref{fig:080905}). The position of
GRB\,090305A, which was previously reported to have no coincident host
to $r\gtrsim 25.6$~mag \citep{ber10}, coincides with an extended NIR
source with $m_{\rm F160W}=25.20 \pm 0.10$~mag which we consider to be
the host galaxy (Figure~\ref{fig:image}; Section~\ref{sec:pcc}). For
GRB\,090426A, we consider the host galaxy to be the source directly
coincident with the afterglow position (Figure~\ref{fig:image};
Section~\ref{sec:pcc}), previously reported to be a ``compact knot''
within a multi-component host galaxy complex from ground-based
observations (\citealt{adp+09,lbb+10}).  For GRB\,091109B, for which a
host galaxy has not been previously reported, the afterglow position is
contaminated by a diffraction spike from a nearby star
(Figure~\ref{fig:himage}). We perform photometry at the afterglow
position using an aperture radius of $2.5\theta_{\rm FWHM}$, and place
a $3\sigma$ limit of $m_{\rm F160W}\gtrsim 25$~mag on a coincident
host galaxy. We note that this NIR limit is substantially shallower
than the $3\sigma$ limit of the image, $\gtrsim 26$~mag. Finally, the
F160W observation of GRB\,130603B has additional flux from a point
source at the optical afterglow position that is not part of the host
galaxy \citep{bfc13,tlf+13}. We perform point spread function (PSF)
subtraction of the point source as described in \citet{bfc13} and
calculate the host photometry from the PSF-subtracted image
(Figure~\ref{fig:image}).

The photometry for all events, as well as $3\sigma$ upper limits for ``host-less'' events are
listed in Table~\ref{tab:hst_obs}. All of our values are consistent
with those  published in the literature from ground-based data
except for GRB\,070714B, where we calculate $m_{\rm F160W} =
22.99 \pm 0.11$~mag while the previously published value is $H = 23.58 \pm
0.20$~mag \citep{gfl+09}.

\tabletypesize{\scriptsize}
\begin{deluxetable*}{lcccccccccccccc}
\tablecolumns{15}
\tablewidth{0pc}
\tablecaption{Short GRB Astrometry and Offsets
\label{tab:offsets}}
\tablehead {
\colhead {GRB}		&
\colhead {Instrument}	&
\colhead {Filter}	&
\colhead {$z$}		&
\colhead {Reference} 		&
\colhead {No.}		&
\colhead {$\sigma_{\rm cat\rightarrow GRB}$} &
\colhead {$\sigma_{\rm GB\rightarrow HST}$} 	&
\colhead {$\sigma_{\rm GRB}$} 		&
\colhead {$\sigma_{\rm gal}$} 		&
\colhead {$\delta$RA}	&
\colhead {$\delta$Dec}	&
\colhead {Offset}	&
\colhead {Offset}	&
\colhead {Offset}       \\
\colhead {}		&
\colhead {}		&
\colhead {}		&
\colhead {}		&
\colhead {}		&
\colhead {}		&
\colhead {(mas)}	&
\colhead {(mas)}	&
\colhead {(mas)}	&
\colhead {(mas)}	&
\colhead {($''$)}	&
\colhead {($''$)}	&
\colhead {($''$)}	&
\colhead {(kpc)}        &
\colhead {($r_e$)}	
}
\startdata
061201  & VLT/FORS2 & $I$     & $0.111?$ & 2MASS & 17 & 115  & \nod & 17 \\
G1      & WFC3      & F160W   &          & VLT   & 14 & \nod & 25   & \nod & 1 & $-11.82$ & $11.14$ & $16.25 \pm 0.03$ & $32.47 \pm 0.06$ & $14.91 \pm 0.03$ \\
G2      & WFC3      & F160W   &          & VLT   & 14 & \nod & 25   & \nod & 2 & $1.69$   & $-0.62$ & $1.80 \pm 0.03$  & $14.47 \pm 0.24^{a}$ & $6.43 \pm 0.11$ \\
\\
070429B & WFC3      & F160W   & $0.902$ & USNO-B & 4  & 150 & \nod  & 1460 & 1.0 & \nod    & \nod    & $<1.46$   & $<11.41$ & $<2.25$ \\
        & WFC3      & F475W   &         & USNO-B & 6  & 290 & \nod  & 1460 & 1.0 & \nod    & \nod    & $<1.46$   & $<11.41$ & \nod$^{b}$ \\
\\
070707  & VLT/FORS1 & $R$     & $<3.6$  & 2MASS & 50  & 156 & \nod  & 10  & \\
        & WFC3      & F160W   &         & VLT   & 76  & \nod & 26 & \nod & 4.3 & $-0.16$ & $-0.01$ & $0.40 \pm 0.03$ & $3.22 \pm 0.24^{a}$ & $1.11 \pm 0.08$ \\
        & ACS       & F606W   &         & VLT   & 97  & \nod & 24 & \nod & 12.5 & $-0.16$ & $0.05$ & $0.40 \pm 0.03$ & $3.22 \pm 0.24^{a}$ & \nod$^{b}$ \\
\\
070714B & Swift/UVOT & $white$ & $0.923$ & 2MASS & 30  & 199 & \nod  & 13 \\
        & WFC3      & F160W   &         & Swift  & 6   & \nod & 110  & \nod & 1 & $-0.40$ & $-1.50$ & $1.55 \pm 0.11$ & $12.21 \pm 0.87$ & $4.56 \pm 0.33$ \\
        & UVIS      & F475W   &         & Swift  & 5   & \nod & 66 & \nod & 7 & $-0.40$ & $-1.50$ & $1.55 \pm 0.07$ & $12.21 \pm 0.53$ & $5.55 \pm 0.24$ \\
\\
070724A & Gemini-N/GMOS & $i$ & $0.4571$ & 2MASS & 6   & 81  & \nod & 13 \\
        & WFC3      & F160W   &          & Gemini & 32  & \nod   & 22 & \nod & 1 & $-0.48$ & $0.81$ & $0.94 \pm 0.03$ & $5.46 \pm 0.14$ & $1.50 \pm 0.04$ \\
\\
070809  & WFC3      & F160W   & $0.473?$ & USNO-B & 5  & 234 & \nod & 400  & 1 & $-5.18$ & $-2.21$ & $5.63 \pm 0.46^{c}$ & $33.22 \pm 2.71$ & $9.25 \pm 0.75$ \\
        & ACS       & F606W   &          & 2MASS & 7   & 220 & \nod & 400  & 2 & $-5.47$ & $-2.10$ & $5.86 \pm 0.46^{c}$ & $35.58 \pm 2.71$ & $9.29 \pm 0.71$ \\
\\
071227  & VLT/FORS2 & $R$     & $0.381$  & 2MASS & 11  & 214 & \nod & 22 \\
        & WFC3      & F160W   &          & VLT   & 11  & \nod & 40 & \nod & 6 & $-2.11$ & $2.11$ & $2.98 \pm 0.05$ & $15.50 \pm 0.24$ & $3.28 \pm 0.05$ \\
        & UVIS      & F438W   &          & VLT   & 10  & \nod & 43 & \nod & 4 & $-2.02$ & $1.83$ & $2.72 \pm 0.02$ & $14.18 \pm 0.25$ & \nod$^{b}$ \\  
\\      
080503  & Gemini-N/GMOS & $r$ & \nod     & 2MASS & 22  & 139 & \nod & 10  \\
        & WFC3      & F160W   &          & Gemini  & 58  & \nod & 29 & \nod & 9  & $0.72$ & $-0.53$ & $0.90 \pm 0.03^{c}$ & $7.24 \pm 0.24^{a}$ & $3.46 \pm 0.12$ \\ 
\\
080905A & VLT/FORS2 & $R$     & $0.1218$ & 2MASS & 116 & 159 & \nod  & 78 \\
G1      & WFC3      & F160W   &          & VLT   & 46  & \nod  & 41 & \nod & 1 & $4.34$ & $-7.06$ & $8.29 \pm 0.08$ & $17.96 \pm 0.19$ & $10.36 \pm 0.10$ \\
G1      & WFC3/UVIS & F814W   &          & VLT   & 61  & \nod  & 32 & \nod & 1 & $4.30$ & $-7.07$ & $8.28 \pm 0.08$ & $17.92 \pm 0.19$ & $10.35 \pm 0.10$ \\
G1      & WFC3/UVIS & F606W   &          & VLT   & 102 & \nod  & 34 & \nod & 1 & $4.34$ & $-7.07$ & $8.30 \pm 0.09$ & $17.97 \pm 0.18$ & $10.38 \pm 0.11$ \\
G2      & WFC3      & F160W   &          & VLT   & 46  & \nod  & 41 & \nod & 13 & $0.07$ & $-0.69$ & $0.69 \pm 0.09$ & $5.55 \pm 0.72^{a}$ & $3.45 \pm 0.45$ \\
G2      & WFC3/UVIS & F606W   &          & VLT   & 102 & \nod  & 34 & \nod & 7 & $0.07$ & $-0.79$ & $0.79 \pm 0.09$ & $6.35 \pm 0.72^{a}$  & $3.97 \pm 0.43$ \\
\\
090305A & Gemini-S/GMOS & $r$ & $<4.1$   & 2MASS & 122 & 198 & \nod  & 6 \\
        & WFC3      & F160W   &          & Gemini  & 115 & \nod & 31 & \nod & 5 & $0.09$ & $-0.42$ & $0.43 \pm 0.03$ & $3.46 \pm 0.24^{a}$ & $1.19 \pm 0.08$ \\
\\
090426  & VLT/FORS2 & $R$     & $2.609$  & SDSS & 60 & 120 & \nod & 10  \\
        & WFC3      & F160W   &          & VLT   & 9   & \nod & 29 & \nod & 4  & $-0.04$ & $0.04$ & $0.06 \pm 0.03^{d}$ & $0.45 \pm 0.25$ & $0.29 \pm 0.14$ \\
\\
090510  & WFC3      & F160W   & $0.903$ & USNO-B & 7 & 370 & \nod  & 200  & 2 & $1.24$ & $0.45$ & $1.33 \pm 0.37$ & $10.37 \pm 2.89$ & $1.99 \pm 0.39^e$ \\
\\
090515  & Gemini-N/GMOS & $r$ & $0.403?$ & SDSS & 109 & 160 & \nod & 15 \\
        & WFC3      & F160W   &         & Gemini   & 61 & \nod & 22 & \nod & 1 & $-3.73$ & $13.47$ & $13.98 \pm 0.03^{c}$ & $75.03 \pm 0.15$ & $15.53 \pm 0.03^e$ \\
\\
091109B & VLT/FORS2 & $R$     & \nod    & 2MASS & 20 & 154 & \nod & 28 \\
        & WFC3      & F160W   &         & VLT   & 39 &     & 21 & \nod & 1 & $9.53$ & $6.79$ & $11.70 \pm 0.03^{c}$ & $94.07 \pm 0.24$ & \nod \\
\\
100117A & Gemini-N/GMOS & $r$ & $0.915$ & SDSS & 95 & 154 & \nod & 26 \\
        & WFC3      & F160W   &         & Gemini  & 21 & \nod & 33 & \nod & 1 & $0.16$ & $0.03$ & $0.17 \pm 0.04$ & $1.32 \pm 0.33$ & $0.57 \pm 0.13^e$   \\
\\
130603B & Magellan/IMACS & $r$ & $0.3564$ & SDSS & 17 & 85 & \nod & 10 \\
        & WFC3      & F160W   &         & Magellan & 12 & \nod & 34 & \nod & 1 & $0.93$ & $0.48$ & $1.05 \pm 0.04$ & $5.21 \pm 0.17$ & $1.05 \pm 0.04^e$ \\
        & ACS       & F606W   &         & Magellan & 9  & \nod & 34 & \nod & 1 & $0.93$ & $0.58$ & $1.09 \pm 0.04$ & $5.41 \pm 0.17$ & $1.36 \pm 0.05^e$
\enddata
\tablecomments{$^{a}$ We assume $z=1$ to calculate these projected physical offsets. \\
$^{b}$ The $r_e$ measurement is highly uncertain due to the low signal-to-noise ratio of the observation. \\
$^{c}$ Offsets are calculated for the galaxy with the lowest probability of chance coincidence. \\
$^{d}$ The offset is calculated relative to the source in direct coincidence with the burst position (Figure~\ref{fig:image}). \\
$^{e}$ Effective radii $r_e$ representative of the combined inner and outer S\'{e}rsic components are used to compute offsets: GRB\,090510: $0.95''$, GRB\,090515: $0.9''$, GRB\,100117A: $0.3''$, GRB\,130603B/F160W: $1''$ and GRB\,130603B/F606W: $0.8''$. \\
 }
\end{deluxetable*}

\subsection{Absolute Astrometry}
\label{sec:astrom}

To determine the position of each short GRB afterglow, we perform
absolute astrometry using point sources in common between the
afterglow discovery images and source catalogs (2MASS, SDSS, or USNO-B
depending on availability). If the position of the afterglow is
contaminated by host galaxy light in the discovery image, we perform
image subtraction using the ISIS package \citep{ala00} relative to
late-time observations when the afterglow contribution is
negligible. We then use {\tt SExtractor}\footnotemark\footnotetext{\tt
http://sextractor.sourceforge.net/} to determine the afterglow
position in the subtracted image. To determine the astrometric tie
from the ground-based image to the catalog, we use the IRAF astrometry
routine {\tt ccmap} and find that a second-order polynomial with six
free parameters corresponding to a shift, scale, and rotation in each
coordinate, provides a robust tie in all cases with an average
$\sigma_{\rm cat \rightarrow GRB}\approx 160$ mas. Our afterglow
positions are consistent with published positions in all relevant
cases, albeit with higher precision. In the cases of GRBs\,070809 and
090510, the afterglow discovery images are not available to us so
we use the published positions and uncertainties in our analysis
\citep{ptc+07,nkk+12}. The absolute afterglow positions and
uncertainties are listed in Table~\ref{tab:hst_obs}.

\subsection{Relative Astrometry and Offsets}

To determine the position of each GRB relative to its host galaxy, and
thus measure precise offsets, we perform relative astrometry by
aligning each of the \hst\ observations to the afterglow discovery
images. We consider three sources of uncertainty: the afterglow
position ($\sigma_{\rm GRB}$), the astrometric tie uncertainty between
the ground-based and \hst\ images ($\sigma_{\rm GB \rightarrow HST}$),
and the host galaxy position ($\sigma_{\rm gal}$).

We measure $\sigma_{\rm GRB}$ from each afterglow image, where the
centroiding accuracy depends on the size of the PSF and the
signal-to-noise ratio of the afterglow detection using {\tt
SExtractor}, and find values of $\sigma_{\rm GRB}\approx 10-80$ mas
(Table~\ref{tab:offsets}), except in the case of GRB\,070429B which
has an uncertainty of $1.5''$ ($1\sigma$) from the \swift/XRT
detection of the afterglow. The second source of uncertainty is the
astrometric tie between the afterglow and host galaxy \hst\ images
($\sigma_{\rm GB \rightarrow HST}$), which is determined using the
same method described in Section~\ref{sec:astrom}. We use a range of
$5-120$ common point sources, depending on the depth of the image and
source density and find values of $\sigma_{\rm GB \rightarrow
HST}\approx 20-110$~mas. The number of astrometric tie objects and
resulting RMS values are listed in Table~\ref{tab:offsets}. The final
source of uncertainty is the centroiding accuracy of the host in the
{\it HST} images. To determine this uncertainty we again use {\tt
SExtractor}, and find values of $\sigma_{\rm gal} \approx
1-13$~mas. This is generally the smallest source of uncertainty
(Table~\ref{tab:offsets}).

For each galaxy/filter combination, we use the afterglow and host
position to measure angular offsets, and for the galaxies with known
redshifts we also calculate physical offsets
(Table~\ref{tab:offsets}). We assume $z=1$ for host galaxies without
known redshift, taking advantage of the relatively flat value of the
angular diameter distance at $z \gtrsim 0.5$. Finally, we use the
effective radii, $r_e$, determined from surface brightness profile
fits (Section~\ref{sec:sbfit} and Table~\ref{tab:morph}) to calculate
host-normalized offsets. The offsets and accompanying combined
uncertainties are listed in Table~\ref{tab:offsets}. For GRBs\,070809
and 090510, where we do not have the afterglow discovery images, we
use the published uncertainties of $0.4''$ and $0.2''$, respectively
\citep{ptc+07,nkk+12}, which dominate over all other sources of
uncertainty.

\tabletypesize{\footnotesize}
\begin{deluxetable}{lccccc}
\tablecolumns{6}
\tabcolsep0.0in\footnotesize
\tablewidth{0pc}
\tablecaption{Short GRB Host Galaxy Morphological Properties
\label{tab:morph}}
\tablehead {
\colhead {GRB}		&
\colhead {Filter}	&
\colhead {$n$}          &
\colhead {$r_e$}        &
\colhead {$r_e$}        &
\colhead {$\mu_e$}      \\
\colhead {}		&
\colhead {}	&
\colhead {}          &
\colhead {($''$)}        &
\colhead {(kpc)}        &
\colhead {(AB mag arcsec$^{-2}$)}             	
}
\startdata
061201/G1 & 160W & $1.03$ & $1.09$ & $2.18$ & $21.86$ \\
061201/G2 & 160W & $0.73$ & $0.28$ & $2.25^{a}$ & $25.34$ \\
070429B & 160W & $2.15$ & $0.65$ & $5.08$ & $23.61$ \\
        & 475W & $1.54$ & $0.08^{b}$ & $0.62$ & $24.69$ \\
070707  & 160W & $0.97$ & $0.36$ & $2.89^{a}$   & $27.01$ \\
070714B & 160W & $0.76$ & $0.34$ & $2.68$ & $24.35$ \\
        & 475W & $1.18$ & $0.28$ & $2.20$ & $27.17$ \\
070724A & 160W & $0.92$ & $0.63$ & $3.64$ & $22.16$ \\
070809  & 160W & $3.03$ & $0.61$ & $3.59$ & $21.47$ \\
        & 606W & $3.38$ & $0.65$ & $3.83$ & $23.64$ \\
071227  & 160W & $1.05$ & $0.91$ & $4.72$ & $21.64$ \\
080503  & 160W & $0.32$ & $0.26$ & $2.09^{a}$   & $26.81$ \\
080905A/G1$^{c}$ & 160W & $\approx 1$ & $\approx 1.8$ & $\approx 3.9$ & \nod \\
080905A/G2 & 160W & $0.70$ & $0.20$ & $1.60^{a}$   & $25.72$ \\
090305A & 160W & $0.57$ & $0.36$ & $2.89^{a}$  & $25.96$ \\
090426A & 160W & $0.89$ & $0.21$ & $1.70$ & $25.78$ \\
090510 ($a<0.4''$)  & 160W & $1.27$ & $0.93$ & $7.27$ & $24.92$ \\
090510 ($a>0.4''$) & 160W & $0.44$ & $0.74$ & $5.79$ & $24.26$ \\
090515 ($a<0.85''$) & 160W & $2.95$ & $0.79$ & $4.24$ & $22.02$ \\
090515 ($a>0.85''$)  & 160W & $0.73$ & $1.19$ & $6.39$ & $22.47$ \\
100117A ($a<0.6''$) & 160W & $0.86$ & $0.28$ & $2.20$ & $22.04$ \\
100117A ($a>0.6''$) & 160W & $4.95$ & $0.07$ & $0.55$ & $18.66$ \\
130603B ($a<1''$) & 160W & $0.96$ & $0.62$ & $3.07$ & $22.51$ \\
130603B ($a>1''$) & 160W & $3.81$ & $3.25$ & $2.02$ & $25.71$ \\
130603B ($a<0.2''$) & 606W & $1.98$ & $0.79$ & $3.92$ & $24.97$ \\
130603B ($a>0.2''$) & 606W & $1.29$ & $0.68$ & $3.37$ & $24.29$ 
\enddata
\tablecomments{$^{a}$ Calculated assuming $z=1$. \\
$^{b}$ Due to the low signal-to-noise ratio of this observation, this measurement likely corresponds to a smaller region within the galaxy, and not the entire galaxy itself. \\
$^{c}$ Although we cannot perform a surface brightness profile fit for GRB\,080905A/``G1'', these parameters are estimated from the apparent morphology and effective radius of the bulge component.
}
\end{deluxetable}

\subsection{Surface Brightness Profile Fitting}
\label{sec:sbfit}

\begin{figure*}
\centering
\includegraphics*[angle=0,width=6.4in]{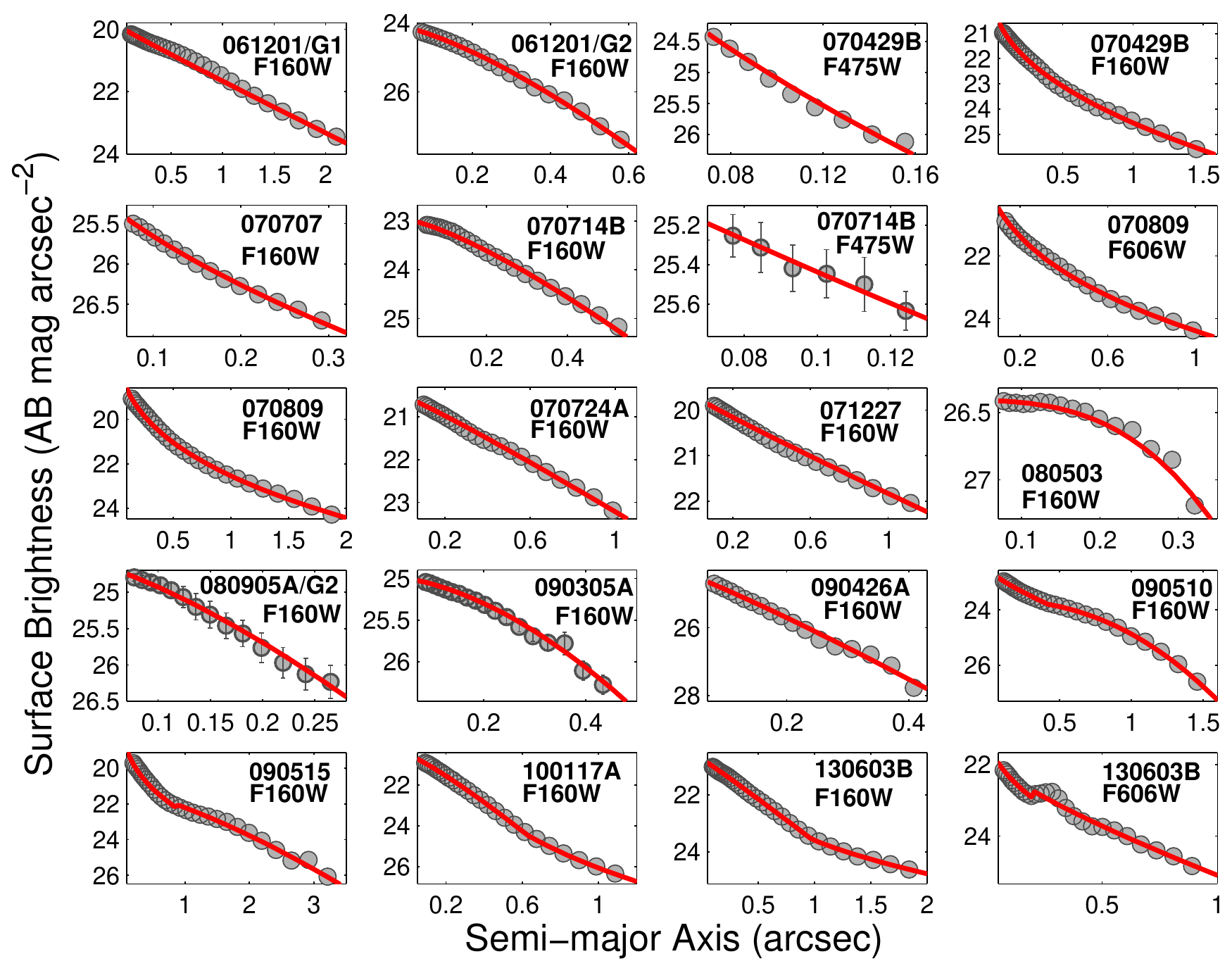}
\caption{Radial surface brightness profiles as determined
from IRAF/{\tt ellipse} (grey circles) and the corresponding best-fit S\'{e}rsic
models (red line) for $16$ host galaxies. For GRBs\,070809,
080503, and 090515, the surface brightness profile for the most
probable host galaxy is shown while for GRB\,061201, the profiles for
both ``G1'' and ``G2'' are shown. For GRB\,080905A, the profile is
shown for the extended source closest to the afterglow position
(Source ``G2'' in Figure~\ref{fig:image}).
\label{fig:sb}}
\end{figure*}

We use the IRAF/{\tt ellipse} routine to generate elliptical intensity
isophotes and construct one-dimensional radial surface brightness
profiles for each galaxy/filter combination. For each observation, we
allow the center, ellipticity, and position angle of each isophote to
vary. In two cases (GRB\,070707/F606W and GRB\,071227/F438W), the
isophotal fit does not converge, which can be attributed to the low
signal-to-noise ratio of these observations. The surface brightness profiles
are displayed in Figure~\ref{fig:sb}.

Using a $\chi^2$-minimization grid search, we fit each profile with a
S\'{e}rsic model given by

\begin{equation}
\Sigma(r)=\Sigma_e\,{\rm exp}\{-\kappa_n[(r/r_e)^{1/n}-1]\},
\label{eqn:sersic}
\end{equation}

\noindent where $n$ is the concentration parameter ($n=1$ is
equivalent to an exponential disk profile, while $n=4$ is the de
Vaucouleurs profile typical of elliptical galaxies), $\kappa_n\approx
2n-1/3+4/405n+46/25515n^2$ is a constant that depends on $n$
\citep{cb99}, $r_e$ is the effective radius, and $\Sigma_e$ is the
effective surface brightness in flux units. We convert $\Sigma_e$ to
units of mag arcsec$^{-2}$, designated as $\mu_e$. In our grid search,
$n$, $r_e$, and $\mu_e$ are the three free parameters. A single
S\'{e}rsic component provides an adequate fit ($\chi^{2}_{\nu} \approx
0.4-1.5$) for most of the host galaxies. In four cases (GRBs\,090510,
090515, 100117A, and 130603B) a single component fit yields $\chi^{2}_{\nu}
\gtrsim 2$. To improve the fit for these cases, we use two separate
S\'{e}rsic components corresponding to the inner and outer regions of
each galaxy. The resulting values for $n$, $r_e$ and $\mu_e$ are
listed in Table~\ref{tab:morph}. We do not perform a surface brightness fit for
GRB\,080905A/``G1'' since it is contaminated by saturated foreground
sources (Figure~\ref{fig:080905}), but given the prominent spiral
structure, the morphology of this host is likely characterized by a
disk profile with $n \sim 1$, and we estimate the size of the
bulge component to be $\approx 1.8''$ ($\approx 3.9$~kpc). The surface
brightness profiles and resulting models are shown in
Figure~\ref{fig:sb}.

\subsection{Host Light Distributions}

\tabletypesize{\footnotesize}
\begin{deluxetable}{lccc}
\tablecolumns{4}
\tabcolsep0.0in\footnotesize
\tablewidth{0pc}
\tablecaption{Short GRB Fractional Flux
\label{tab:lightfrac}}
\tablehead {
\colhead {GRB}		&
\colhead {Instrument}	&
\colhead {Filter}	&
\colhead {$1\sigma$ Fractional Flux}	
}
\startdata
050709$^{a}$ & WFPC2 & F450W & 0  \\
        & ACS   & F814W & 0.09 \\
\\
050724$^{a}$ & WFPC2 & F450W & 0.03 \\
       & WFPC2 & F814W & 0.33 \\
\\
051221$^{a}$ & WFPC2 & F555W & 0.54 \\
       & WFPC2 & F814W & 0.65 \\
\\
060121$^{a}$ & ACS   & F606W & 0.41 \\
\\
060313$^{a}$ & ACS   & F475W & 0.04 \\
       & ACS   & F775W & 0    \\
\\
061006$^{a}$ & WFPC2 & F555W & 0.56 \\
       & ACS   & F814W & 0.63 \\
\\
061201$^{ab}$ & ACS  & F606W & 0 \\
             & ACS  & F814W & 0 \\
             & WFC3/IR & F160W & 0 \\
\\
070707  & WFC3/IR   & F160W & 0     \\
        & ACS       & F606W & 0     \\
\\
070714B & WFC3/IR   & F160W & 0 \\
        & WFC3/UVIS & F475W & 0     \\
\\
070724A & WFC3/IR   & F160W & 0.23 \\
\\
070809$^{b}$  & WFC3/IR   & F160W & 0     \\
        & ACS       & F606W & 0     \\
\\ 
071227  & WFC3/IR   & F160W & 0.19 \\
        & WFC3/UVIS & F438W & 0     \\
\\
080503$^{b}$  & WFC3/IR   & F160W & 0     \\
\\
080905A & WFC3/IR   & F160W & 0 \\
        & WFC3/UVIS   & F814W & 0 \\
        & WFC3/UVIS   & F606W & 0 \\
\\
090305A & WFC3/IR   & F160W & 0.30 \\
\\
090426  & WFC3/IR   & F160W & 0.82 \\
\\
090510  & WFC3/IR   & F160W & 0     \\
\\
090515$^{b}$  & WFC3/IR   & F160W & 0     \\
\\
100117A & WFC3/IR   & F160W & 0.54 \\
\\
130603B & WFC3/IR & F160W & 0.27 \\
        & ACS/606W & F606W & 0.35
\enddata
\tablecomments{Fraction of host galaxy light in pixels fainter than
the GRB position. \\
$^{a}$ From \citet{fbf10}. \\
$^{b}$ No coincident host to the depth of available \hst\ imaging.}
\end{deluxetable}

To determine the brightness of the galaxy at the GRB location relative
to the host light distribution, we follow the methodology of
\citet{fls+06}, \citet{kkp08}, and \citet{fbf10}, and calculate for
each galaxy image the fraction of total light in pixels fainter than
the afterglow position (``fractional flux'';
Table~\ref{tab:lightfrac}). Eleven bursts have differential
astrometric positions of better than one pixel
(Table~\ref{tab:offsets}). If the afterglow position spans multiple
pixels, we take the average brightness among those pixels to be the
representative flux of the afterglow position. For each image, we
create an intensity histogram of a region centered on the host galaxy
and determine a $1\sigma$ cut-off level for the host by fitting a
Gaussian profile to the sky brightness distribution (equivalent to a
signal-to-noise ratio cut-off of 1). We then plot the pixel flux
distribution above the appropriate cut-off level for a region
surrounding the host, and determine the fraction of light in pixels
fainter than the afterglow pixel (see Table~\ref{tab:lightfrac}).

For GRBs\,071227/F160W and 130603B/F606W, we mask sources
contaminating the position of the galaxy and set these pixels to the
brightness level of the surrounding pixels. We note that for
GRB\,080905A, we are unable to explicitly calculate the fractional
flux with respect to ``G1'' due to the presence of saturated stars, while for GRB\,061201 we cannot
make a unique host association (Section~\ref{sec:pcc}). However, both
of these bursts have afterglow locations at the level of the
sky background, and thus have fractional flux values of
zero regardless of their host associations. Table~\ref{tab:lightfrac} also
includes the values for seven bursts from \citet{fbf10}.

\section{Analysis}

\label{sec:analysis}

\subsection{Probabilities of Chance Coincidence}
\label{sec:pcc}

To assess the probability that each of the bursts originated from a
coincident galaxy or from another galaxy in the field, we perform
aperture photmetry for galaxies within the \hst\ field of view, discarding
noticeably fainter galaxies with increasing distance from the burst
since these objects will have a lower probability of being the host
galaxy.  We then calculate the probability of chance coincidence,
$P_{cc}(<\delta R)$ for each galaxy based on the distance from the
burst position ($\delta R$) and apparent magnitude ($m$)
(c.f., \citealt{bkd02,ber10}). For bursts at offsets of $<1r_e$ (GRBs\,090426
and 100117A), we use the effective size of the galaxy, $\delta R =
r_e$, while for the remaining cases, we take $\delta R$ to be the
projected distance between the burst and candidate host center. For
the bursts with previously established hosts (Figure~\ref{fig:image}),
we find $P_{cc}(<\delta R) \approx 10^{-4}-10^{-3}$, consistent
with ground-based results \citep{fbc+13}; the next probable hosts
have values at least one order of magnitude greater, with
$P_{cc}(<\delta R) \approx 0.02-0.30$. The lowest value of $0.02$ is
for GRB\,090426A, which has two galaxies within $1''$ of the optical
afterglow position, in addition to the source at the GRB
position. From previous ground-based observations, these sources were
considered to comprise a single host galaxy complex
\citep{aap+09,lbb+10}. However, given the lack of apparent interaction
between these sources in the \hst\ image (see Figure~\ref{fig:image})
and the comparatively low value of $P_{cc}(<\delta R) \approx 10^{-4}$
for the source at the afterglow position, we consider the latter to be the host galaxy. For
GRB\,080905A/``G1'' (Figure~\ref{fig:080905}), we are unable to accurately
determine the brightness in any of the filters due to the presence of
foreground saturated stars, but using $R\sim 18$ mag determined from
\citet{rwl+10} and our offset of $8.3''$, we find that ``G1''
has $P_{cc}(<\delta R) \approx 0.01$, while ``G2'' has
$P_{cc}(<\delta R) \approx 0.08$. Therefore, ``G1'' is the
more likely host galaxy, although this case is less clear than the
other previously established host associations at small offsets.

For bursts with no obvious host galaxy at small offsets from previous
ground-based or \hst\ observations (Figure~\ref{fig:himage}),
termed ``host-less'' (GRBs\,061201, 070809, 080503, 090305A, 090515:
\citealt{ber10}; 091109B: \citealt{gcn10154}), we calculate a range
of probabilities, $P_{cc}(<\delta R) \approx 6 \times
10^{-3}-0.08$. The associations are most robust for GRBs\,070809
and 090305A with the most probable hosts at offsets of $5.7''$
($P_{cc}(<\delta R) \approx 6 \times 10^{-3}$) and $0.43''$
($P_{cc}(<\delta R) \approx 7 \times 10^{-3}$), respectively. The host
association for GRB\,070809 is the same as that made in \citet{ber10},
although here we calculate a lower probability of chance coincidence
by a factor of three. We note that the extended source we associate
with GRB\,090305A was not detected in ground-based observations to
$r\gtrsim 25.6$~mag \citep{ber10}. Due to the low probability of
chance coincidence and the lack of more likely host galaxy candidates in
the \hst\ observation, we consider this to be the host galaxy. We
calculate moderate probabilities of $\approx 0.05$ for GRBs\,080503
and 090515, with the most probable hosts at offsets of $0.90''$ and
$14''$, respectively. These associations are the same as those
previously published \citep{ber10,pmg+09} but the probabilities of
chance coincidence are lower by a factor of two in this work.

The associations are more ambiguous for GRBs\,061201 and 091109B. For
GRB\,061201, the two most probable host galaxies (``G1'' and ``G2'';
Figure~\ref{fig:himage}) have offsets of $16.3''$ and $1.8''$,
respectively, and both have $P_{cc}(<\delta R) \approx 0.07$ so the
host association is inconclusive. However, ``G1'' is at a relatively
low redshift of $z=0.111$ \citep{sdp+07} while ``G2'' is likely at
$z\approx 1$; thus the physical offsets are $\gtrsim 15$~kpc in both
cases (Table~\ref{tab:offsets}). For GRB\,091109B, the position is
contaminated by a diffraction spike, but if there is no coincident
host galaxy at $\gtrsim 25$~mag, the most probable host has $m_{\rm
F160W} \approx 19.8$~mag at an offset of $11.7''$, yielding
$P_{cc}(<\delta R) \approx 0.08$. We note that \hst\ imaging at a
different rotation angle will be essential in determining whether this
burst originates from a host galaxy at $\lesssim 1''$ separation or from a galaxy
at a larger offset. Due to the uncertainty of the associations for
GRBs\,061201 and 091109B, we do not include these bursts in our
subsequent offset analysis.

Overall, these probability of chance coincidence results agree with
those in the literature \citep{pmg+09,ber10,rwl+10}, and provide deep
NIR limits of $\gtrsim 26.2$~mag for bursts which lack hosts at
$\delta R \lesssim$few arcsec. We discuss the possibility that such
bursts originated from galaxies fainter than the detection threshold
of the \hst\ observations (and demonstrate that this is unlikely) in
more detail in Section~\ref{sec:disc}.

\subsection{Morphological Properties}

Using the results from the radial surface brightness profiles
(Figure~\ref{fig:sb}), we classify the short GRB hosts in terms of
their morphological parameters: S\'{e}rsic value, $n$, and effective
size, $r_e$. We find two elliptical galaxies, the hosts of GRBs\,070809 and 090515,
with $n \approx 3.0-3.4$ while the remaining galaxies have disk-like
morphologies with $n \approx 0.3-2.1$ (Table~\ref{tab:morph}). We note
that GRB\,100117A exhibits a complex morphology in the NIR, with
S\'{e}rsic indices of $n\approx 0.9$ and $\approx 5$ for its inner and
outer regions, respectively, although it is spectroscopically
classified as an early-type galaxy with a stellar population age of
$\approx 1-2$~Gyr and no evidence for star formation activity \citep{fbc+11}. GRB\,130603B, which is a
star-forming galaxy with SFR$\gtrsim 1.3\,M_{\odot}$~yr$^{-1}$
\citep{cpp+13}, has an inner component in the NIR with
$n \approx 1$ and a broad outer component with $n \approx 3.8$. This
host galaxy has an irregular, asymmetric morphology in the optical
band with excess flux at radial distances of $a \approx 0.2-0.4''$
in the surface brightness profile (Figure~\ref{fig:sb}) and S\'{e}rsic
components with $n\approx 2$ and $1.3$.

The effective radii range from $\approx 0.2-1.2''$ with a median size
of $0.36''$. We note that the signal-to-noise ratio of the
GRB\,070429B/F475W observation is low and the $r_e$ measurement likely
corresponds to a smaller region within the galaxy, and not the entire
galaxy. For the host galaxies that require two S\'{e}rsic
components, we use the radius which encloses half of the flux as the
effective size when computing the host-normalized offsets. These
values are $0.95''$ (GRB\,090510), $0.9''$ (GRB\,090515), $0.3''$
(GRB\,100117A), $1''$ (GRB\,130603B/F160W) and $0.8''$
(GRB\,130603B/F606W). For the short GRBs with known redshifts, the median
physical size is about $r_e \approx 3.6$~kpc. The smallest hosts are GRBs\,070714B and
100117A while the largest are GRBs\,090510 and 090515. The median
value for this sample is the same as the value of $3.5$~kpc reported
in \citet{fbf10} for a preliminary sample of hosts. Compared to the
long GRB median host galaxy size of $1.7$~kpc \citep{wbp07}, short GRB
host galaxies are twice as large. This is consistent with their larger
luminosities \citep{ber09} and stellar masses \citep{lb10}.

\subsection{Offsets}

\begin{figure}
\includegraphics*[angle=0,width=3.4in]{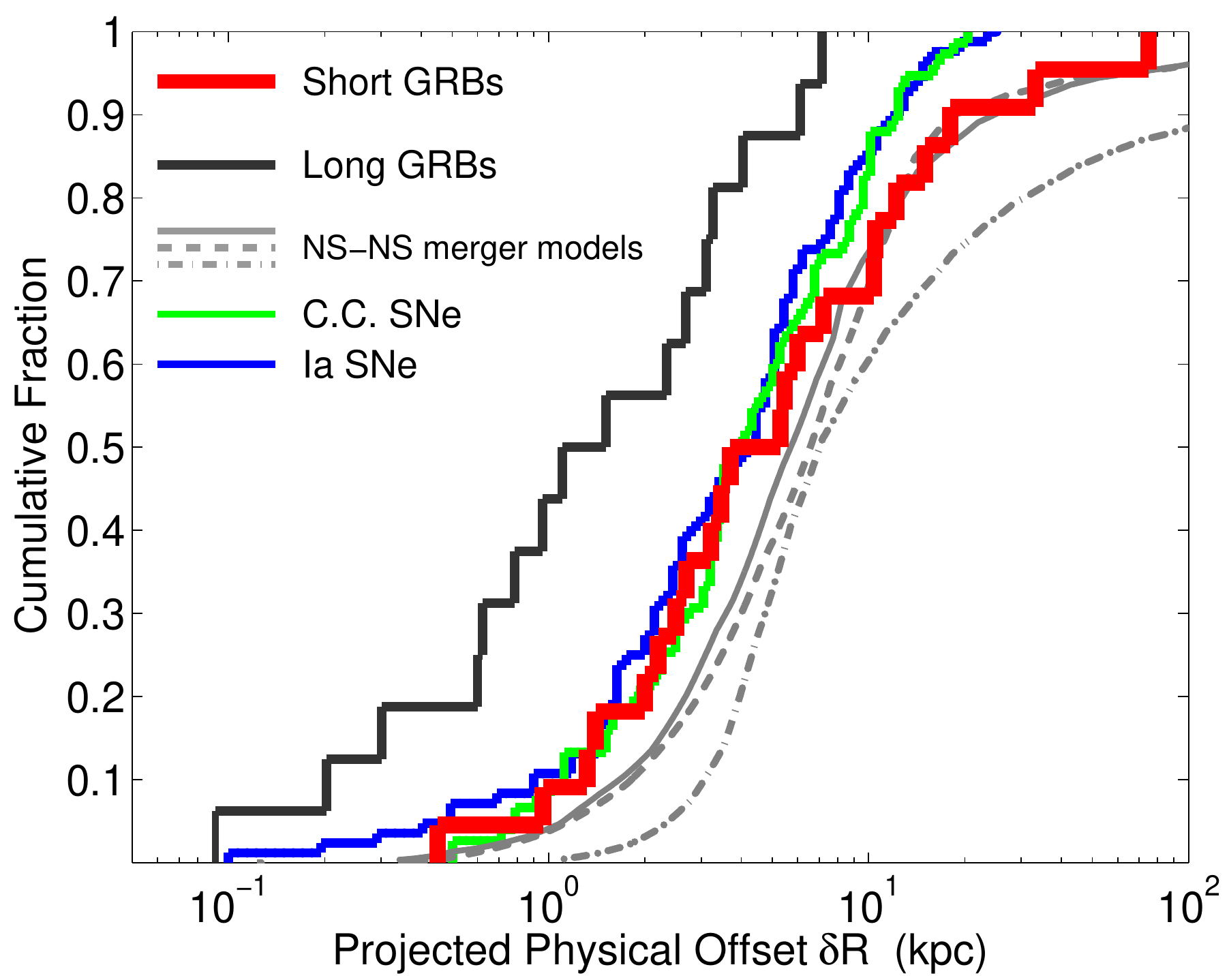}
\caption{Cumulative distribution of projected physical offsets for 22
short GRBs with sub-arcsecond positions (red; \citealt{fbf10}, this
work, and 3 ground-based measurements:
\citealt{fbm+12,mbf+12,sta+12,bzl+13}). For five bursts with no
spectroscopic redshifts (GRBs\,060121, 070707, 080503, 090305A, and
111020A), we have assumed $z=1$. Also shown are the cumulative
distributions for long GRBs (black; \citealt{bkd02}), core-collapse
SNe (green; \citealt{psb08}), Type Ia SNe (blue;
\citealt{psb08}), and predicted offsets for NS-NS binaries (grey;
\citealt{fwh99,bsp99,bpb+06}).
\label{fig:offset}}
\end{figure}

\begin{figure*}
\begin{tabular}{cc}
\includegraphics*[angle=0,width=3.4in]{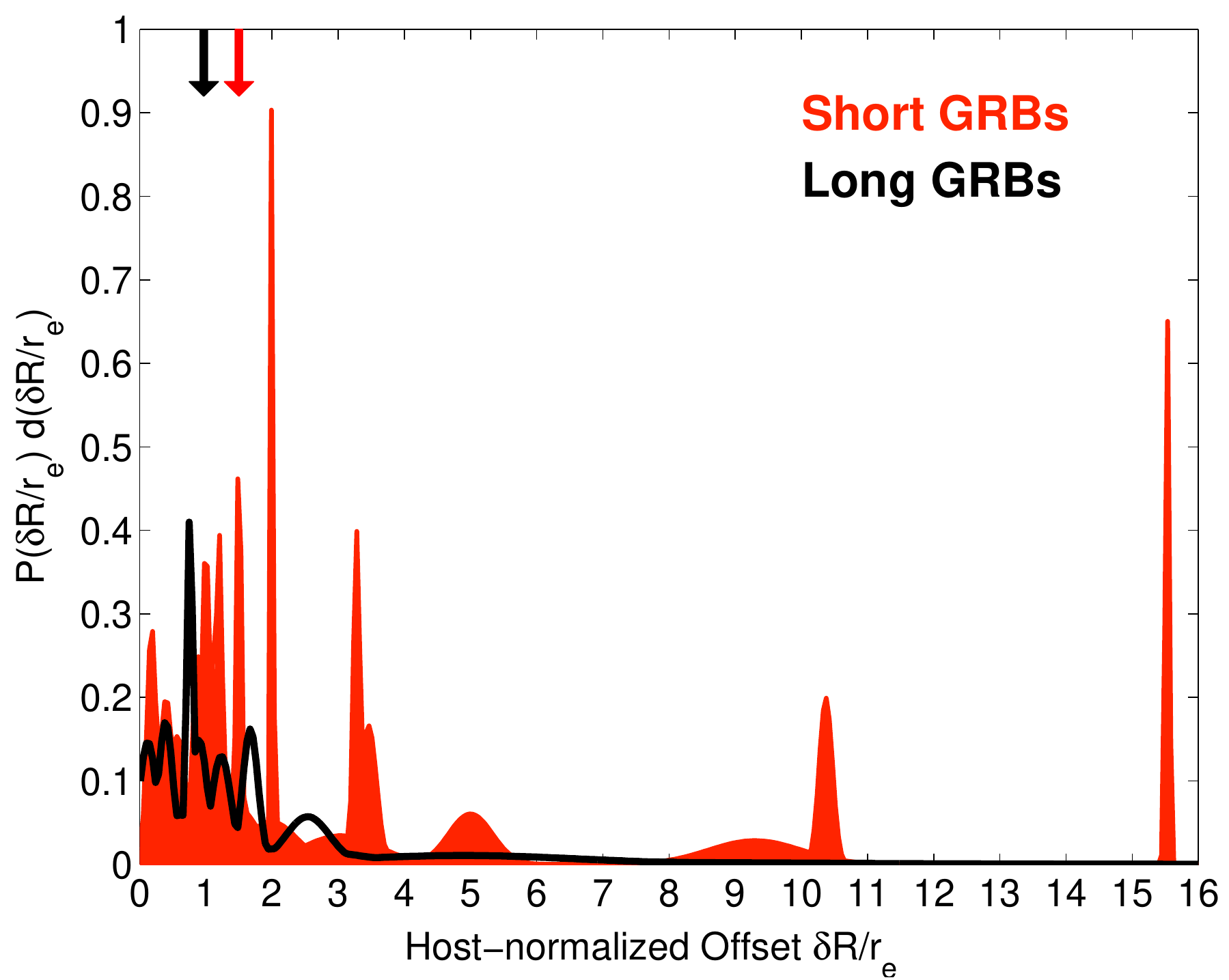} &
\includegraphics*[angle=0,width=3.4in]{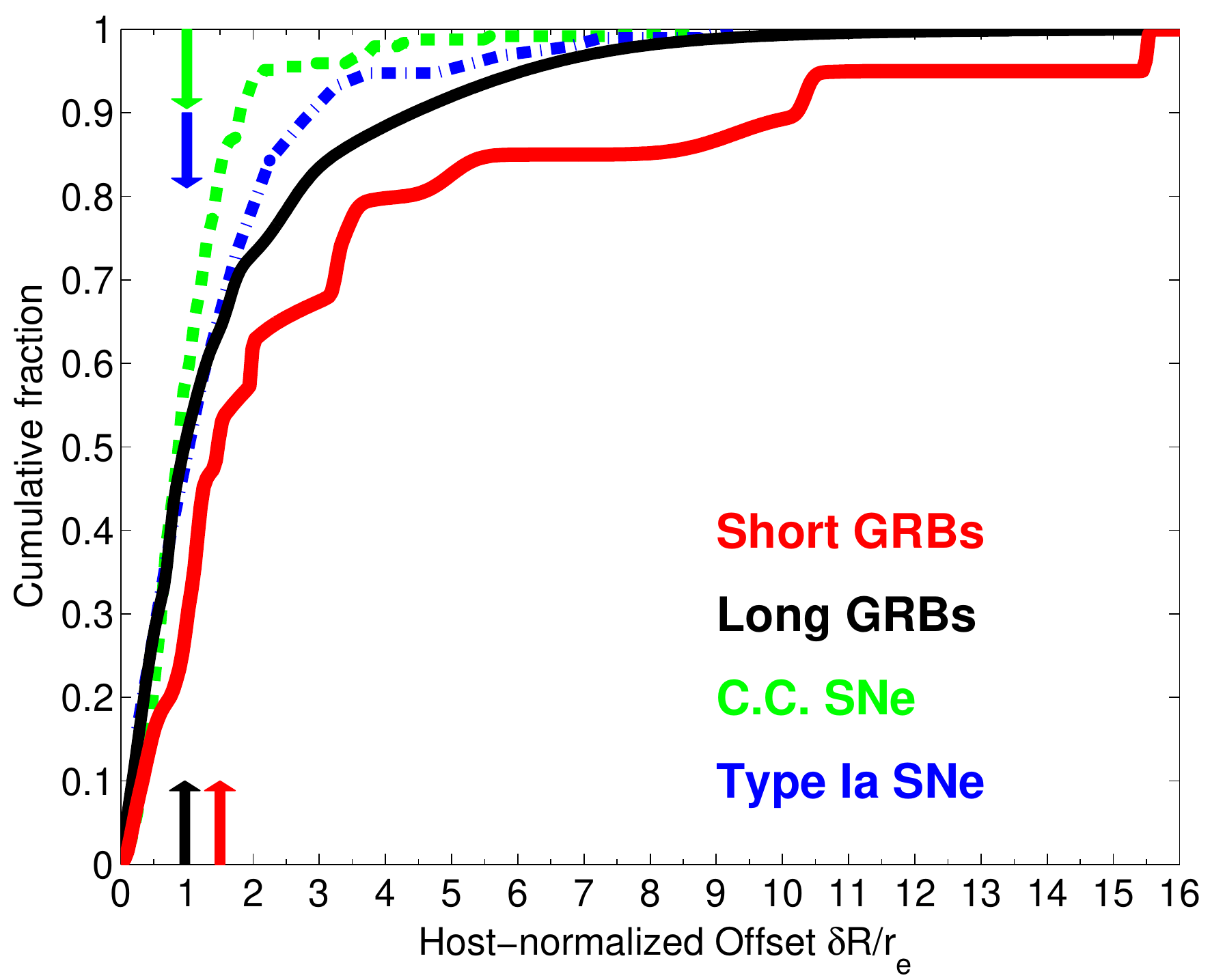}
\end{tabular}
\caption{{\it Left}: Differential distributions of host-normalized
offsets in units of effective radius, $r_e$, accounting for the
uncertainty in each offset measurement, for short GRBs (red shaded region) and long GRBs
(black line). The sample is comprised of $20$ short GRBs with resolved host galaxies
from {\it HST} data (\citealt{fbf10} and
this work), including GRB\,050509B which
has only an XRT position. Arrows denote the weighted median offset for
each population: $1.0 r_e$ (long) and $1.5 r_e$ (short). {\it
Right}: Cumulative host-normalized offset distributions for short GRBs
(red) and long GRBs (black). Also shown are the distributions for core-collapse
supernovae (green dashed; \citealt{kk12}) and Type Ia supernovae (blue
dot-dashed; \citealt{gmo+12}). Arrows denote the weighted median offset for
each population; the median for SNe is also $\approx 1r_e$.
\label{fig:offset2}}
\end{figure*}

To study the locations of short GRBs with respect to their host
galaxies, we first consider the distribution of projected angular
offsets. The range of angular offsets is $\approx 0.1-14''$ with
GRB\,090426 as the smallest
offset\footnotemark\footnotetext{Calculated from the galaxy in direct
coincidence with the optical afterglow position, using $z=2.609$ as
determined from afterglow spectroscopy \citep{adp+09,lbb+10}.} and
GRB\,090515 as the largest. From the angular offsets, we calculate
projected physical offsets, assuming $z \approx 1$ for bursts without known
redshift. We find a range of $\approx 0.5-75$~kpc (Figure~\ref{fig:offset2}).

We supplement this sample of offsets with six measurements from
\citet{fbf10}. In addition, we use offset measurements from
ground-based observations of all of the remaining short GRBs with
sub-arcsecond positions: GRB\,111020A with $6 \pm 1$~kpc (assuming
$z \approx 1$, \citealt{fbm+12}), GRB\,111117A with $10.5 \pm 1.7$~kpc
\citep{mbf+12,sta+12}, and GRB\,120804A with $2.2 \pm 1.2$~kpc
\citep{bzl+13}. Therefore, the full sample of offsets includes $22$ short
GRBs (Figure~\ref{fig:offset}) with a resulting median offset of
$4.5$~kpc. In comparison to the long GRB median offset of $1.3$~kpc,
the short GRB median offset is $\approx 3.5$ times larger. The short
GRB median offset is comparable to those for Type Ia and core-collapse
SNe of $\approx 3$~kpc (Figure~\ref{fig:offset}; \citealt{psb08}), but
the short GRB offset distribution extends to much larger offsets: only
$10\%$ of both SN types have offsets of $\gtrsim 10$~kpc, compared to
$25\%$ for short GRBs. Furthermore, no SNe have offsets of $\gtrsim
20$~kpc, while $10\%$ of short GRBs do.

In Figure~\ref{fig:offset}, we also show a comparison of the short GRB offset
distribution to the predicted distributions from population
synthesis models of NS-NS binary mergers in Milky Way type galaxies
\citep{fwh99,bsp99,bpb+06}. The short GRB distribution is broadly
consistent with the NS-NS binary merger predictions, and is in very good
agreement with two of the three models \citep{bsp99,bpb+06}. The median offset
for the predicted distributions is $5-7$~kpc, slightly larger than the
observed median of $4.5$~kpc. We note that the observed distribution
is mainly derived from short GRBs with optical afterglows and may be
missing a few bursts with less precise localization from X-ray
afterglows \citep{fbc+13} that may occur outside of their host
galaxies. Thus, while the observed distribution of offsets should be
fairly representative of the true distribution, accounting for such
missing events would only extend the distribution to larger offsets,
in even better agreement with the NS-NS merger models.

A study by \citet{tko+08} suggested that short GRBs with extended
emission in the X-rays have smaller offsets than short GRBs with no
such emission. Two bursts in our sample, GRBs\,070714B and 080503 have
reported evidence for extended emission at $\gtrsim 5\sigma$
significance \citep{gcnr70,gcnr138,pmg+09}. GRB\,070714B has an offset of
$\approx 12.2$~kpc while GRB\,080503 has an offset of $\approx
7.2$~kpc from its most probable host assuming $z \approx 1$. Combining
these two bursts with four bursts analyzed in \citet{fbf10} with
sub-arcsecond positions and extended emission (GRBs\,050709, 050724,
061006, and 060121), the median offset for the population is
$3.2$~kpc, with a range of $\approx 1-12$~kpc. For the remaining $16$ bursts
with no extended emission and precise offset measurements, the median
offset is $5.3$~kpc. A Kolmogorov-Smirnov (K-S) test
comparing the two populations gives a $p$-value of $0.9$, supporting
the null hypothesis that the two populations are drawn from the same
underlying distribution. Therefore, there is no clear evidence from
their locations that short GRBs with and without extended emission
require different progenitor systems.

To compare the offset distributions in a more uniform manner, we
calculate host-normalized offsets, $\delta R/r_e$, using the effective
radii as determined from our morphological analysis
(Section~\ref{sec:sbfit}). We find a range of host-normalized offsets
of $\approx 0.3-15.5~r_e$ for the bursts with sub-arcsecond positions
(Table~\ref{tab:offsets}). We supplement this sample with seven
measurements from \citet{fbf10}, one of which has only an XRT position
and thus a more uncertain offset (GRB\,050509B). To account for the
varying uncertainty in each offset, we plot a differential
distribution of host-normalized offsets following the methodology of
\citet{bkd02}, as well as the resulting cumulative distribution
(Figure~\ref{fig:offset2}). The total sample of short GRBs with
host-normalized offsets is comprised of $20$~events, with a median of
$\approx 1.5~r_e$ and only about $25\%$ of the events at $\lesssim 1r_e$. For comparison, the host-normalized offset
distributions for long GRBs \citep{fls+06}, core-collapse SNe
\citep{kkp08} and Type Ia SNe \citep{gmo+12} have median offsets of
$\approx 1~r_e$. Furthermore, a K-S test comparing the host-normalized
offsets for long and short GRBs does not support the null hypothesis
that the two populations are drawn from the same underlying
distribution ($p=0.03$). A K-S test between short GRBs and Type Ia SNe
yields $p=10^{-3}$, indicating that the two populations are drawn from
different host-normalized offset distributions. Indeed, $\approx 20\%$
of short GRBs have offsets of $\gtrsim 5r_e$, compared to only
$\approx 5\%$ for Type Ia SNe.

\subsection{Light Fraction}

\tabletypesize{\footnotesize}
\begin{deluxetable*}{lccccccc}
\tablecolumns{8}
\tabcolsep0.0in\footnotesize
\tablewidth{0pc}
\tablecaption{Fractional Flux Statistics
\label{tab:ffstat}}
\tablehead {
\colhead {} & 
\colhead {} &
\colhead {} &
\colhead {} &
\multicolumn{4}{c}{K-S test $p$-values} \\
\cline{5-8} \\
\colhead {Sample}		&	
\colhead {Band}	&
\colhead {Median} &
\colhead {Percentage at zero$^{a}$} &
\colhead {Linear$^{b}$} & 
\colhead {Long GRBs} &
\colhead {C. C. SNe} &
\colhead {Type Ia SNe}
}
\startdata
Short GRBs (all)$^{c}$ & Optical & 0.15 & 45 & 0.04 & \nod  & \nod & 0.02 \\

Short GRBs ($P_{cc}(<\delta R)\lesssim 0.01$) & Optical  & 0.25 & 29 & 0.07 & \nod & \nod & 0.16 \\

Short GRBs       & UV      & 0 & 55 & 0.01 & 0.002 & 0.002 & 0.34 \\

Long GRBs & UV & 0.83 & 6 & 0.01 & \nod & 0.001 & $3.8 \times 10^{-4}$ \\

C. C. SNe & UV & 0.60 & 5 & 0.42 & 0.001 & \nod & 0.04 \\

Type Ia SNe & Optical & 0.34 & 6 & 0.35 & \nod & \nod & \nod \\

Type Ia SNe & UV & 0.35 & 34 & 0.08 & $3.8 \times 10^{-4}$ & 0.04 & \nod
\enddata
\tablecomments{
$^{a}$ Percentage of a given population with fractional flux values of zero. \\
$^{b}$ Corresponds to a distribution that linearly tracks host galaxy light. \\
$^{c}$ All $20$ short GRBs with sub-arcsecond positions. \\
 }
\end{deluxetable*}

\begin{figure*}
\centering
\begin{tabular}{cc}
\includegraphics*[angle=0,width=3.4in]{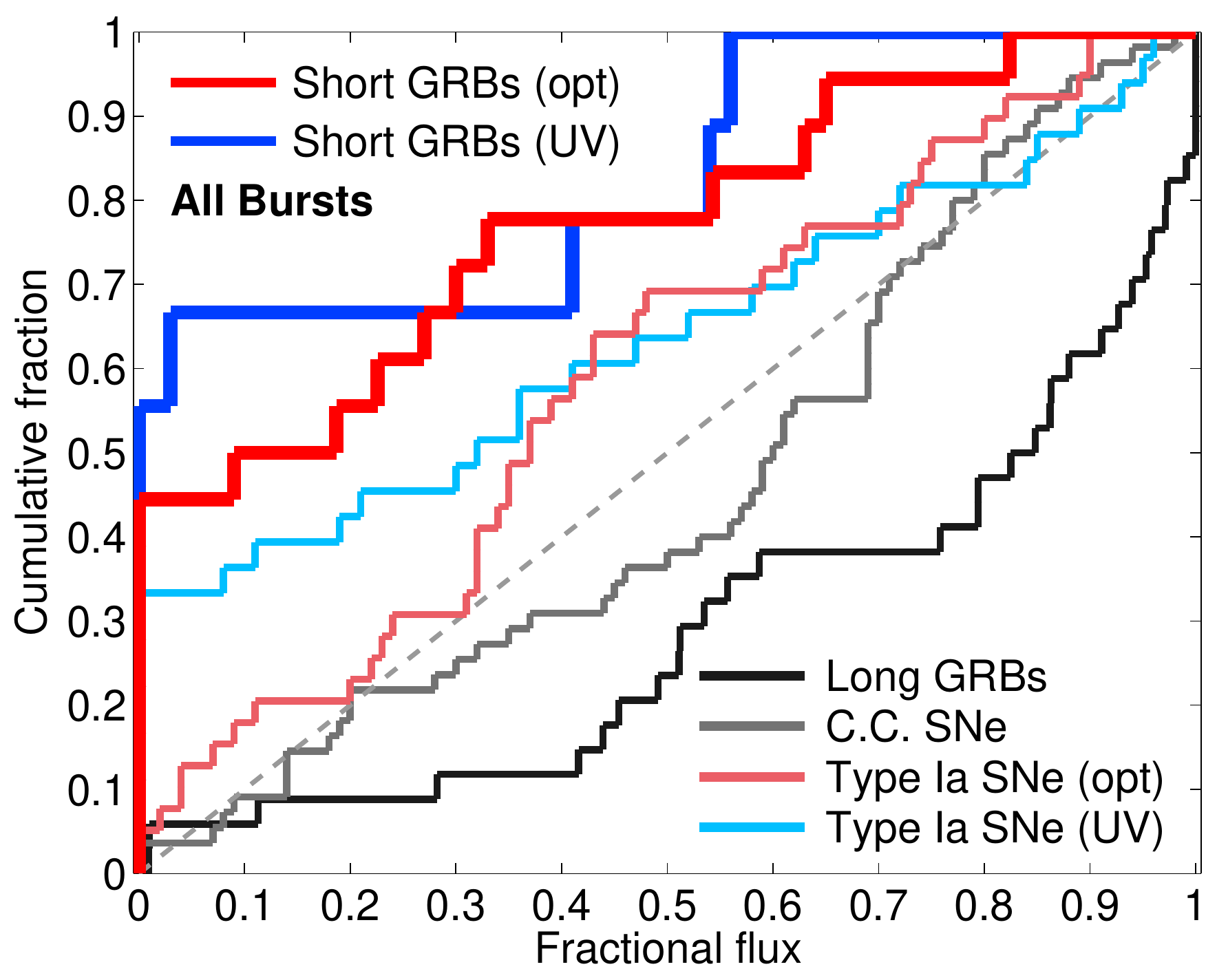} 
\includegraphics*[angle=0,width=3.4in]{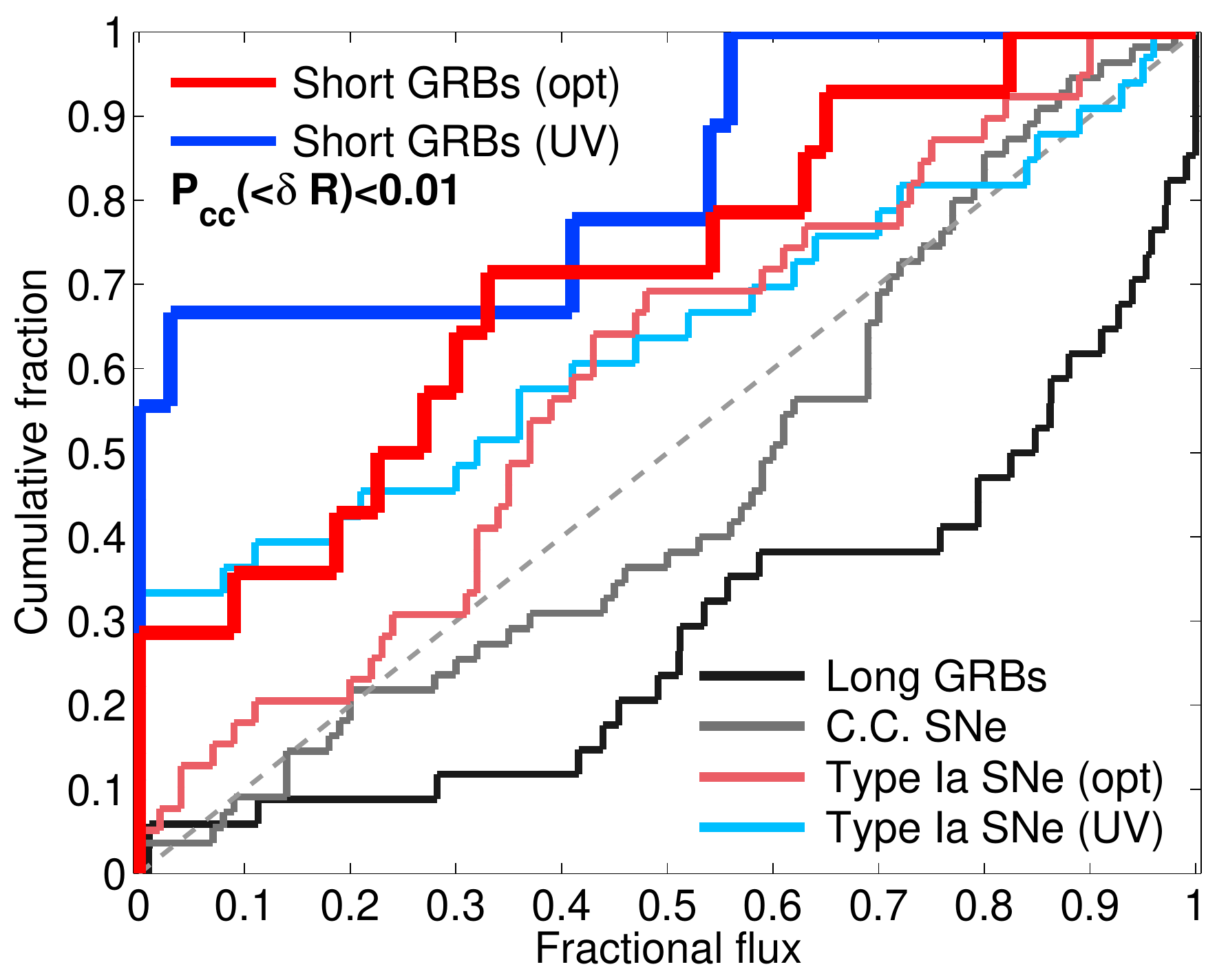}  
\end{tabular}
\caption{{\it Left:} Cumulative distribution of the fractional flux at
short GRB locations relative to their host light. The rest-frame
optical ($0.4-1.2\,\mu$m; red) and rest-frame UV ($<0.4\,\mu$m; blue)
for $20$ short GRBs with \hst\ observations are shown (\citealt{fbf10}
and Table~\ref{tab:lightfrac} in this work). For the purpose of
determining the rest-frame band, bursts without known redshifts are
assigned $z=1$. Also shown are the distributions for
``normal-velocity'' Type Ia supernovae for $u'$-band (light blue) and
$r'$-band (pink; \citealt{wwf+13}), core-collapse supernovae (grey;
\citealt{slt+10}) and long GRBs (black; \citealt{fls+06,slt+10}). {\it
Right:} Cumulative distribution of fractional flux including only
bursts with host associations with $P_{cc}(<\delta R) \lesssim
0.01$. This case excludes four bursts: GRBs\,061201, 080503,
080905A and 090515.
\label{fig:lightfrac}}
\end{figure*}

To further study the local explosion environments of short GRBs,
we utilize the fractional flux method which, unlike the spatial offset
method, is independent of host morphology. We divide the fractional
flux values into two categories based on the bursts' observed filters
and redshifts: rest-frame UV ($\lambda_{\rm rest} \lesssim 0.4$\,$\mu$m) tracking star formation activity and
rest-frame optical ($\lambda_{\rm rest} \gtrsim 0.4$\,$\mu$m) tracking stellar mass. For bursts
with no redshift, we assume fiducial values of $z=1$ to determine the proper rest-frame band. For the $14$
bursts in this analysis, all have rest-frame optical measurements
while only three have rest-frame UV measurements (GRB\,070707 assuming
$z=1$, GRBs\,070714B and 071227; Table~\ref{tab:lightfrac}). Despite
having coincident host galaxies, these three bursts are located on the
lowest level of their hosts' UV light with fractional flux
measurements of zero. The rest-frame optical measurements span a range
of $0-0.8$, with GRB\,090426 as the highest measurement
(Table~\ref{tab:lightfrac}).

We supplement these data with six additional bursts analyzed in
\citet{fbf10}, bringing the total sample size to $20$ events. We find
that the resulting distributions are strongly skewed to low fractional
flux measurements: $45\%$ of short GRBs are located on the lowest optical
brightness level of their hosts (fractional flux $\approx 0$), and
$55\%$ are on the lowest UV level (Figure~\ref{fig:lightfrac} and
Table~\ref{tab:ffstat}). The short GRB distributions have very low
median fractional flux values of $\approx 0.15$ for the optical and
zero for the UV (Table~\ref{tab:ffstat}). Furthermore, $\approx 75\%$
of the events are located on the faint end (fractional flux $\lesssim
0.5$) of their hosts' optical and UV regions
(Figure~\ref{fig:lightfrac}). A K-S test comparing the observed short
GRB distribution to a distribution that is linearly correlated with
host galaxy light (diagonal line in Figure~\ref{fig:lightfrac}) yields
$p$-values of $0.04$ and $0.01$ for the optical and UV, respectively
(Table~\ref{tab:ffstat}). These results demonstrate that short GRBs
are not correlated with their hosts' rest-frame UV and optical light,
i.e., they do not trace regions of star formation or even stellar
mass.

The short GRB distribution is particularly striking when compared to
long GRBs, which lie on the brightest UV regions of their host
galaxies and have a median fractional flux of $0.83$
(Figure~\ref{fig:lightfrac} and Table~\ref{tab:ffstat};
\citealt{fls+06,slt+10}). Core-collapse SNe, which have a median value
of $0.60$, may slightly over-represent their hosts' UV light
\citep{slt+10}, commensurate with their origin in star-forming
regions. Furthermore, at most a few percent of long GRBs and
core-collapse SNe lie on the faintest regions of their host galaxies
(Table~\ref{tab:ffstat}).

On the other hand, Type Ia SNe, which result from older stellar
progenitor systems, slightly under-represent their hosts' UV and
optical light distributions, although K-S test results indicate that
this is only marginal (Table~\ref{tab:ffstat}). In particular, $34\%$
of Type Ia SNe have UV fractional flux values of zero, compared to
$55\%$ for short GRBs. The difference is more apparent in the optical,
with only $6\%$ of the Type Ia SNe population located on the faintest
regions of their hosts, compared to $45\%$ for short GRBs
(Table~\ref{tab:ffstat}). A K-S test comparing the distributions of
short GRBs and Type Ia SNe indicates that the populations are not
drawn from the same underlying distribution in the optical ($p =
0.02$).

We cannot completely rule out the remote possibility that the events
with less robust host associations ($P_{cc}(<\delta R)\gtrsim 0.01$)
and fractional flux values of zero instead originate from faint host
galaxies with $m_{\rm F160W}\gtrsim 26.2$~mag, below the detection
threshold of the \hst\ images. In this scenario, the fractional flux
values for these events may be greater than zero. Therefore, if we
only include events with $P_{cc}(<\delta R)<0.01$ (thereby excluding
GRBs\,061201, 080503, 080905A and 090515 from the distribution), we
find that the short GRBs are still uncorrelated with their hosts'
optical light ($p=0.07$; Table~\ref{tab:ffstat}), with a median value
of $\approx 0.25$ (Figure~\ref{fig:lightfrac}). It is important to
note that even in this conservative case, $\approx 30\%$ of the bursts
lie on the lowest optical flux levels of their hosts
(Figure~\ref{fig:lightfrac}). The short GRB UV distribution is
unaffected since there are no excluded bursts with rest-frame UV
measurements.

\section{Implications for the Progenitors}
\label{sec:disc}

\begin{figure*}
\begin{tabular}{cc}
\includegraphics*[angle=0,width=3.4in]{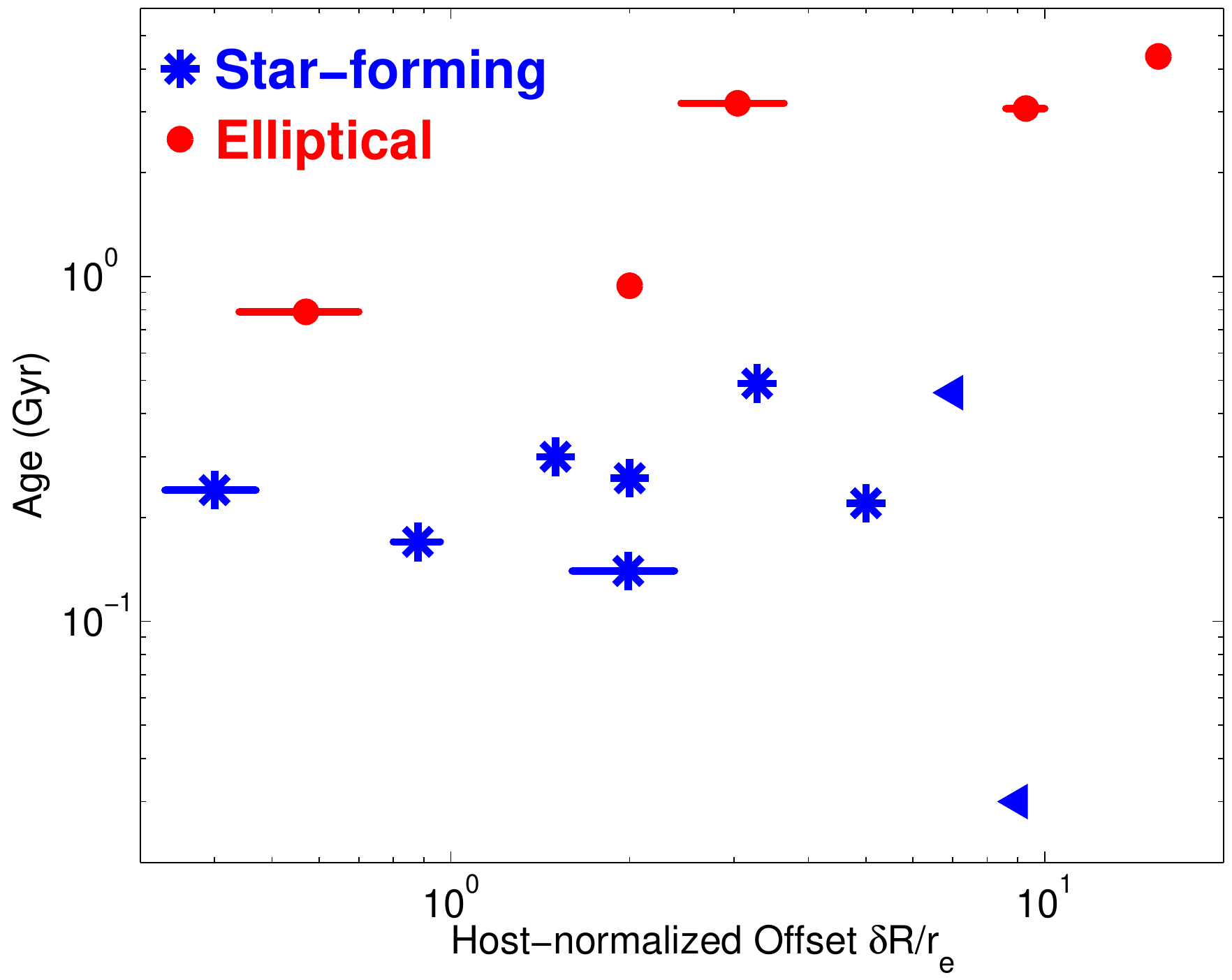} &
\includegraphics*[angle=0,width=3.4in]{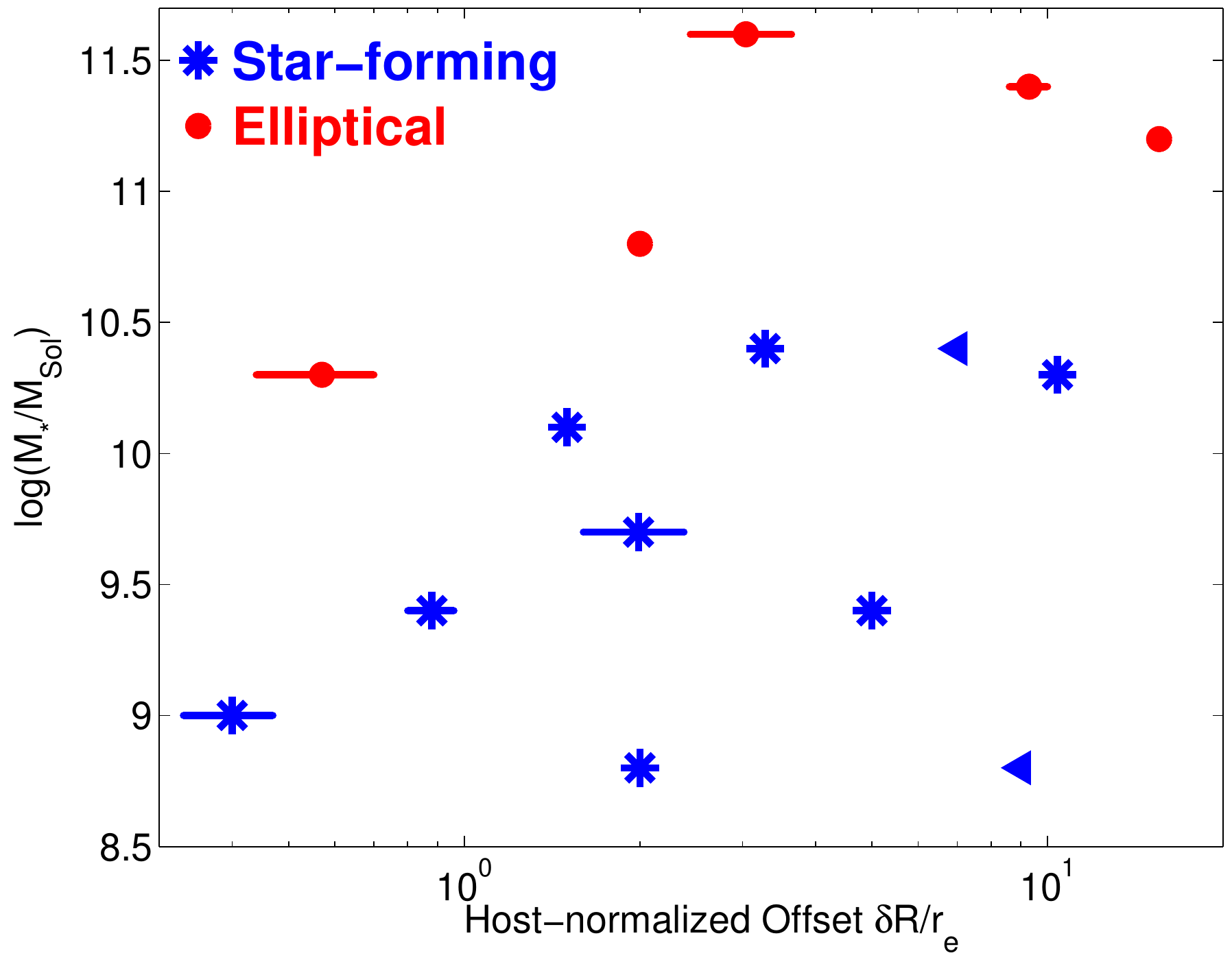}
\end{tabular}
\caption{Stellar population age (left) and stellar mass (right) versus
host-normalized offsets for $15$ short GRB host galaxies
\citep{lb10,rwl+10}, including three bursts with only XRT positions
(GRBs\,050509B, 051210 and 070429B). Star-forming (blue stars) and
elliptical hosts (red circles), as determined from spectroscopy, are
indicated. Triangles denote bursts with no detected optical afterglow
but that have a single galaxy within the XRT error circle and
therefore an upper limit on the offset. We find no obvious trends
between stellar mass and host-normalized offsets, or between stellar
population age and host-normalized offsets.
\label{fig:massage}}
\end{figure*}

\begin{figure}
\centering
\includegraphics*[angle=0,width=3.2in]{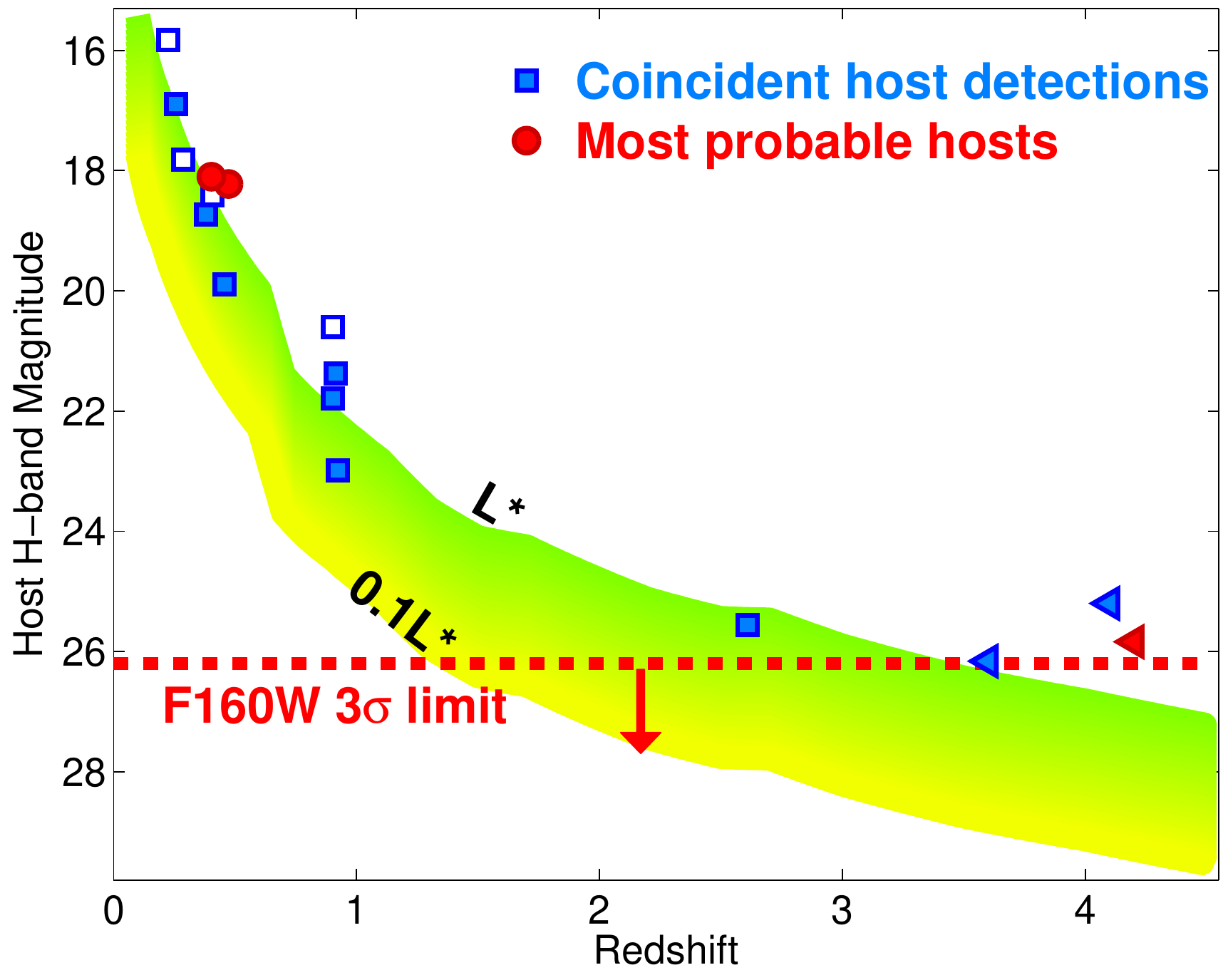}
\caption{$H$-band host galaxy apparent magnitudes from ground-based
 and \hst\ observations versus redshift for short GRBs with coincident
 host galaxies (blue squares and triangles) and the most probable hosts
 of GRBs\,070809, 080503 and 090515 (red circles and
 triangle). Triangles denote upper limits on the burst redshift from
 the detection of the optical afterglow in a particular band. Open
 symbols correspond to bursts with XRT positions only and a probable
 host galaxy (GRBs\,050509B, 060502B, 070429B and 100206A;
 \citealt{bpc+07,pmm+12}). The average $3\sigma$ upper limit of
 $m_{\rm F160W} \approx 26.2$~mag for the bursts lacking coincident
 host galaxies (dotted red line) and the evolving luminosity function
 for $0.1L^{*}$ to $L^{*}$ galaxies (green shaded area;
 \citealt{cmm+03,pgf+03,sfc+06,mvq+07,hdc+10,rpb+11,msb+12,sm13}) are
 shown.
\label{fig:lumz}}
\end{figure}

Using the host associations, morphologies, offset distributions, and
light distributions presented in the previous sections, we draw
implications about the progenitor systems of short GRBs.  We first
note that the morphological analysis for the 16 host galaxies
presented in this paper continues to support the apparent dominance of
late-type host galaxies (star-forming with disk morphologies), with
only $\approx 1/4$ of short GRBs in hosts with elliptical morphologies
\citep{ber09,fbf10,fbc+13}.  The tendency for short GRBs to occur in
star-forming galaxies indicates that the rate of short GRBs is driven
by both recent star formation activity and stellar mass, as also
inferred from the distribution of host galaxy masses \citep{lb10}.  In
addition, the effective sizes of short GRB hosts are significantly
larger than those for long GRB hosts, consistent with their larger
luminosities, stellar masses, and metallicities \citep{ber09,lb10}.

In terms of locations relative to the host centers, we find that short
GRBs span a wide range of projected physical offsets of $\sim 0.5-75$
kpc, with a median value of about 4.5 kpc and with about $25\%$ of all
events occurring at $\gtrsim 10$ kpc.  The median offset is $3.5$
times larger than for long GRBs \citep{bkd02,fbf10}.  The larger
offsets of short GRBs are also evident when normalizing by the
effective radii of their hosts, with $\delta R/r_e\approx 0.3-16$ and
a median value of $\delta R/r_e\approx 1.5$.  In addition, only $25\%$
of short GRBs have offsets of $\lesssim r_e$.  The median value is
$1.5$ times larger than for long GRBs, core-collapse SNe, and even
Type Ia SNe, which have $\langle\delta R/r_e\rangle \approx 1$.  The
broader distribution relative to long GRBs and SNe, and the fact that
only about $20\%$ of short GRBs occur within the radius that contains
half the light indicate that short GRB progenitors migrate from their
birth-sites before producing the bursts.  Taken together with the
overall match to population synthesis predictions
(Figure~\ref{fig:offset}), the physical and host-normalized offsets
point to compact object binary progenitors with significant kicks.

In this context, the observed offset distribution depends on the combined
distributions of kick velocities, merger timescales, and host galaxy
masses.  Therefore, the observed offset distribution, combined with
estimates of the host galaxy stellar masses and merger timescales
\citep{lb10} can provide insight into the kick velocity distribution.
Considering the host stellar population ages as a proxy for the merger
timescale, we expect the bursts with the largest offsets to originate
from elliptical galaxies, since these systems have had the most time
to travel prior to merger.  However, the offsets are also sensitive
to the escape velocities, and thus stellar masses, of the hosts, and
we therefore expect short GRBs in low-mass galaxies to have larger
offsets \citep{rrd03,bpb+06,zrd09}.  We investigate these effects
using the host-normalized offsets from this work and \citet{fbf10} in
conjuncion with inferred stellar masses and stellar population ages
from \citet{lb10}.  In Figure~\ref{fig:massage} we plot the values of
$\delta R/r_e$ versus stellar mass and population age for both early-
and late-type hosts.  We calculate the Kendall~$\tau$ coefficient,
(where a value of $\tau=1$ indicates statistical correlation), to assess
whether the stellar population properties are correlated with
host-normalized offsets, discarding the offset upper limits. We find that $\tau \approx 0.34$ ($p=0.09$)
for stellar mass and $\tau \approx 0.26$ ($p=0.20$) for stellar
population age. Both results agree with the null hypothesis that
there is no strong correlation, and thus we find no clear trend between
stellar population properties and offsets.  In particular, we find
that short GRBs in both elliptical and star-forming galaxies span the
full range of host-normalized offsets.  This result suggests that the
observed offset distribution is primarily determined by the
distribution of kick velocities.

Using the projected physical offsets and the stellar population ages
as a proxy for the merger timescale we can calculate the minimum
projected kick velocities if the progenitors originate at the host centers
($v_{\rm kick,min}$).  We find a range of $v_{\rm kick,min}\approx
2-150$ km s$^{-1}$, with a median of about $16$ km s$^{-1}$.  However,
a more reasonable value for the kick velocity of each system needs to
take into account the host velocity dispersion ($v_{\rm disp}$), and
we therefore use the geometric mean, $\sqrt{v_{\rm kick,min}\,v_{\rm
disp}}$ (c.f., \citealt{bpc+07}).  Using a fiducial value for the
late-type hosts of $v_{\rm disp}\approx 120$ km s$^{-1}$, as measured
for the Milky Way \citep{bhm+05,xrz+08}, and $\approx 250$ km s$^{-1}$
inferred for $\sim 10^{11}$ M$_{\odot}$ elliptical galaxies
\citep{fp99}, we find projected kick velocities of $\approx 20-140$ km
s$^{-1}$ with a median of $\approx 60$~km~s$^{-1}$. This range is
consistent with the inferred natal kick velocities for the eight known
Galactic NS-NS binaries, which range from $\approx 5-500$ km s$^{-1}$
\citep{fk97,wlh+06,wwk10}.

Independent of the offset distribution, the locations of short GRBs
relative to their hosts' light distribution also point to
explosion sites that are distinct from the progenitor birth sites.  In
particular, our analysis clearly demonstrates that short GRBs are not
spatially correlated with either star forming regions or even with the
underlying distribution of stellar mass.  This is unlike long GRBs and
core-collapse SNe, which track UV light \citep{fls+06,slt+10}, or Type
Ia SNe, which generally track stellar mass \citep{kkp08}.  Most strikingly,
about $30-45\%$ of all short GRBs occur in regions that effectively
contain no rest-frame optical light, and hence negligible stellar mass
(Figure~\ref{fig:lightfrac}), indicating that the progenitors were not
formed at the explosion sites.  Moreover, studies of the host galaxy
demographics and stellar mass distribution show that the short GRB
rate depends on both stellar mass and star formation activity
\citep{lb10,fbc+13}, while the light distributions point to explosion
sites that are de-coupled from both star-forming regions and the
stellar mass distribution.  The combination of these results, along
with the large host-normalized offsets, provides the strongest support
to date for NS-NS/NS-BH progenitors with significant migration from
their birth sites to their eventual explosion sites.

Finally, the \hst\ observations presented here provide unprecedented
NIR limits on coincident hosts for short GRBs previously termed as
host-less based on ground-based optical data (GRBs\,061201, 070809,
080503 and 090515).  Using these limits we investigate the possibility
that these events are not at large offsets from their hosts (as
appears to be the case based on probability of chance coincidence
arguments; \citealt{ber10} and this paper), but instead originate from
coincident hosts that are below the \hst\ detection limit.

To determine the combination of luminosity and redshift required for
such faint coincident hosts, we compare the average $3\sigma$ limit of
the \hst\ observation, $m_{\rm 160W}\gtrsim 26.2$ mag, to
the observer-frame $H$-band galaxy luminosity function taking into
account its evolution with redshift (Figure~\ref{fig:lumz}).  We find
that if these hosts are $\sim L^*$ galaxies, typical of other short
GRB hosts (Figure~\ref{fig:lumz}; \citealt{ber09}), they would need to
originate at $z\gtrsim 3.5$.  The highest known short GRB redshift is
$z=2.609$ (GRB\,090426), while typical redshifts are $\sim 0.2-1$, so
in this scenario, these bursts would represent a distinct population
of the highest redshift short GRBs.  If instead the bursts have
redshifts following the observed redshift distribution, this would
require the hosts to have luminosities well below $0.1\,L^{*}$
(Figure~\ref{fig:lumz}), at least an order of magnitude below the
typical luminosities of short GRB hosts.  Furthermore, \citet{ber10}
show that these bursts have systematically fainter optical afterglows
than bursts with coincident hosts, which is at odds with the scenario
of similar redshifts and sub-luminous hosts.  Thus, we do not consider
the possibility that the host-less events have coincident hosts below
the detection limit of the \hst\ data as likely.  Instead, when
combined with a probability of chance coincidence analysis
(\citealt{ber10} and this paper), these bursts appear to be associated
with galaxies that are typical of the short GRB host population
(Figure~\ref{fig:lumz}), with resulting offsets of $\sim 10-100$ kpc.
Thus, the deep NIR limits presented here provide further evidence for
large offsets consistent with NS-NS/NS-BH progenitors.

\section{Conclusions}

We presented \hst\ observations and a detailed analysis of $22$
short GRB host galaxies. Based on this analysis
combined with the results from \citealt{fbf10}, we draw several key
conclusions:

\begin{enumerate}

\item Short GRB host galaxies with disk morphologies dominate the
sample, with only $\approx 1/4$ of the hosts having elliptical
morphologies. The median effective size of short GRB hosts is $\approx
3.6$~kpc, about twice as large as long GRB hosts, which are
exclusively associated with late-type, star-forming galaxies.

\item Short GRBs have projected physical offsets from their host
galaxies of $\approx 0.5-75$~kpc, with a median of $\approx 4.5$~kpc,
$3.5$ times larger than the offsets for long GRBs. Compared to the
distributions for core-collapse and Type Ia SNe, short GRBs extend to
larger offsets, with $\approx 25\%$ of events at $\gtrsim 10$~kpc,
compared to only $10\%$ for both SN types.

\item Taking into account their host sizes, short GRBs have
host-normalized offsets of $0.3-15.5r_e$ with a median of $\approx
1.5r_e$, approximately $1.5$ times larger than those for long GRBs,
core-collapse SNe, and Type Ia SNe. Furthermore, $\approx 20\%$ of
short GRBs have offsets of $\gtrsim 5r_e$, compared to only $5\%$ for
Type Ia SNe, which also result from old stellar progenitors.

\item In the context of NS-NS/NS-BH progenitors, we use the offset
distribution, stellar population age distribution, and typical
velocity dispersions for star-forming and elliptical galaxies, to
infer kick velocities of
$\approx 20-140$~km~s$^{-1}$ with a median of $\approx 60$~km~s$^{-1}$. This
is generally consistent with the range of kick velocities inferred for
Galactic NS-NS binaries.

\item Short GRBs severely under-represent their hosts' rest-frame UV
or optical light. In particular, $30-45\%$ of short GRBs are located
on the faintest optical regions of their host galaxies, while $\approx
55\%$ occur in the faintest UV regions, showing that short GRBs do not
spatially track star formation or stellar mass. Combined with the host
galaxy demographics which imply a short GRB rate driven by both star
formation and stellar mass, this demonstrates that short GRBs migrate
from their birth sites to their eventual explosion sites and provides
strong support for progenitor kicks, ie., NS-NS/NS-BH mergers.

\item For bursts with no robust association to a host galaxy within
$\sim$few arcsec, we consider a faint coincident host origin by
comparing the NIR limit of $m_{\rm F160W}\gtrsim 26.2$~mag to the
$H$-band galaxy luminosity function and other short GRB hosts. If
these hosts are $\sim L^{*}$ galaxies, typical of other short GRB
hosts, they are constrained to $z\gtrsim 3.5$. Alternatively, if these
bursts occur at typical short GRB redshifts of $z\sim 0.5$, this
requires sub-luminous hosts compared to the population, with $\lesssim
0.1 L^{*}$. Instead, it is more likely that these bursts originate
from $\sim L^{*}$ galaxies at $\approx 10-100$~kpc offsets as
indicated by probability of chance coincidence analysis.

\end{enumerate}

Through this analysis, we have provided independent lines of evidence
which argue for NS-NS/NS-BH mergers as the progenitors of short
GRBs. In particular, both the spatial offsets and their locations on
the faintest regions of their hosts demonstrate that the progenitors
must migrate between their formation and the eventual explosions. In
addition, we have made detailed comparisons between short GRBs and
Type Ia SNe which, unlike long GRBs and core-collapse SNe, result from
old stellar progenitors. We find that the two populations differ in
their distributions of host-normalized offsets and rest-frame optical
light locations, with short GRBs having larger offsets and a weaker
correlation with stellar mass. Finally, we note that the large
fraction of short GRBs with a weak correlation to stellar light agrees
with the overall indication from the afterglow emission that the
parsec-scale densities around the progenitors are generally low,
$\lesssim 0.1$ cm$^{-3}$
\citep{sbk+06}.

\acknowledgments

The Berger GRB group is supported by the
National Science Foundation under Grant AST-1107973, and previously by
NASA/Swift AO6 grant NNX10AI24G and A07 grant NNX12AD69G. This paper
includes data gathered with the 6.5 meter Magellan Telescopes located
at Las Campanas Observatory, Chile.  This work is based in part on
observations obtained at the Gemini Observatory, which is operated by
the Association of Universities for Research in Astronomy, Inc., under
a cooperative agreement with the NSF on behalf of the Gemini
partnership: the National Science Foundation (United States), the
Science and Technology Facilities Council (United Kingdom), the
National Research Council (Canada), CONICYT (Chile), the Australian
Research Council (Australia), Ministério da Ciência, Tecnologia e
Inovação (Brazil) and Ministerio de Ciencia, Tecnología e Innovación
Productiva (Argentina). This work made use of data supplied by the UK
Swift Science Data Centre at the University of Leicester. Based on
observations made with ESO Telescopes at the La Silla Paranal
Observatory under programme ID 59.A-9002(D), 079.D-0909(C),
080.D-0906(G), 081.D-0588(C), 083.D-0606(C), and 084.D-0621(B). Based
on observations made with the NASA/ESA Hubble Space Telescope,
obtained from the data archive at the Space Telescope Science
Institute. STScI is operated by the Association of Universities for
Research in Astronomy, Inc. under NASA contract NAS 5-26555.

\clearpage


\begin{thebibliography}{0}
\expandafter\ifx\csname natexlab\endcsname\relax\def\natexlab#1{#1}\fi
\expandafter\ifx\csname bibnamefont\endcsname\relax
  \def\bibnamefont#1{#1}\fi
\expandafter\ifx\csname bibfnamefont\endcsname\relax
  \def\bibfnamefont#1{#1}\fi
\expandafter\ifx\csname citenamefont\endcsname\relax
  \def\citenamefont#1{#1}\fi
\expandafter\ifx\csname url\endcsname\relax
  \def\url#1{\texttt{#1}}\fi
\expandafter\ifx\csname urlprefix\endcsname\relax\def\urlprefix{URL }\fi
\providecommand{\bibinfo}[2]{#2}
\providecommand{\eprint}[2][]{\url{#2}}

\end{thebibliography}


\begin{thebibliography}{}

\bibitem[\protect\citeauthoryear{{Alard}}{{Alard}}{2000}]{ala00}
{Alard}, C. 2000, \aaps, 144, 363

\bibitem[\protect\citeauthoryear{{Antonelli} et~al.}{{Antonelli}
  et~al.}{2009a}]{adp+09}
{Antonelli}, L.~A., et~al. 2009a, \aap, 507, L45

\bibitem[\protect\citeauthoryear{{Antonelli} et~al.}{{Antonelli}
  et~al.}{2009b}]{aap+09}
{Antonelli}, L.~A., et~al. 2009b, \aap, 507, L45

\bibitem[\protect\citeauthoryear{{Battaglia} et~al.}{{Battaglia}
  et~al.}{2005}]{bhm+05}
{Battaglia}, G., et~al. 2005, \mnras, 364, 433

\bibitem[\protect\citeauthoryear{{Belczynski} et~al.}{{Belczynski}
  et~al.}{2006}]{bpb+06}
{Belczynski}, K., {Perna}, R., {Bulik}, T., {Kalogera}, V., {Ivanova}, N.,  \&
  {Lamb}, D.~Q. 2006, \apj, 648, 1110

\bibitem[\protect\citeauthoryear{{Berger}}{{Berger}}{2009}]{ber09}
{Berger}, E. 2009, \apj, 690, 231

\bibitem[\protect\citeauthoryear{{Berger}}{{Berger}}{2010}]{ber10}
{Berger}, E. 2010, \apj, 722, 1946

\bibitem[\protect\citeauthoryear{{Berger} et~al.}{{Berger}
  et~al.}{2009}]{bcf+09}
{Berger}, E., {Cenko}, S.~B., {Fox}, D.~B.,  \& {Cucchiara}, A. 2009, \apj,
  704, 877

\bibitem[\protect\citeauthoryear{{Berger}, {Fong}, \& {Chornock}}{{Berger}
  et~al.}{2013}]{bfc13}
{Berger}, E., {Fong}, W.,  \& {Chornock}, R. 2013, ArXiv e-prints

\bibitem[\protect\citeauthoryear{{Berger} et~al.}{{Berger}
  et~al.}{2013}]{bzl+13}
{Berger}, E., et~al. 2013, \apj, 765, 121

\bibitem[\protect\citeauthoryear{{Bloom}, {Kulkarni}, \& {Djorgovski}}{{Bloom}
  et~al.}{2002}]{bkd02}
{Bloom}, J.~S., {Kulkarni}, S.~R.,  \& {Djorgovski}, S.~G. 2002, \aj, 123, 1111

\bibitem[\protect\citeauthoryear{{Bloom} et~al.}{{Bloom} et~al.}{2007}]{bpc+07}
{Bloom}, J.~S., et~al. 2007, \apj, 654, 878

\bibitem[\protect\citeauthoryear{{Bloom}, {Sigurdsson}, \& {Pols}}{{Bloom}
  et~al.}{1999}]{bsp99}
{Bloom}, J.~S., {Sigurdsson}, S.,  \& {Pols}, O.~R. 1999, \mnras, 305, 763

\bibitem[\protect\citeauthoryear{{Cenko} et~al.}{{Cenko} et~al.}{2008}]{cbn+08}
{Cenko}, S.~B., et~al. 2008, ArXiv e-prints

\bibitem[\protect\citeauthoryear{{Chapman} et~al.}{{Chapman}
  et~al.}{2008}]{clw+08}
{Chapman}, R., {Levan}, A.~J., {Wynn}, G.~A., {Davies}, M.~B., {King}, A.~R.,
  {Priddey}, R.~S.,  \& {Tanvir}, N.~R. 2008, in American Institute of Physics
  Conference Series, Vol. 983, 40 Years of Pulsars: Millisecond Pulsars,
  Magnetars and More, ed. {C.~Bassa, Z.~Wang, A.~Cumming, \& V.~M.~Kaspi}, 301

\bibitem[\protect\citeauthoryear{{Chen} et~al.}{{Chen} et~al.}{2003}]{cmm+03}
{Chen}, H.-W., et~al. 2003, \apj, 586, 745

\bibitem[\protect\citeauthoryear{{Ciotti} \& {Bertin}}{{Ciotti} \&
  {Bertin}}{1999}]{cb99}
{Ciotti}, L.,  \& {Bertin}, G. 1999, \aap, 352, 447

\bibitem[\protect\citeauthoryear{{Cucchiara} et~al.}{{Cucchiara}
  et~al.}{2013}]{cpp+13}
{Cucchiara}, A., {Prochaska}, J.~X., {Perley}, D.~A., {Cenko}, S.~B., {Werk},
  J., {Cao}, Y., {Bloom}, J.~S.,  \& {Cobb}, B.~E. 2013, ArXiv e-prints

\bibitem[\protect\citeauthoryear{{D'Avanzo} et~al.}{{D'Avanzo}
  et~al.}{2009}]{dmc+09}
{D'Avanzo}, P., et~al. 2009, \aap, 498, 711

\bibitem[\protect\citeauthoryear{{Eichler} et~al.}{{Eichler}
  et~al.}{1989}]{elp+89}
{Eichler}, D., {Livio}, M., {Piran}, T.,  \& {Schramm}, D.~N. 1989, \nat, 340,
  126

\bibitem[\protect\citeauthoryear{{Evans} et~al.}{{Evans} et~al.}{2009}]{ebp+09}
{Evans}, P.~A., et~al. 2009, \mnras, 397, 1177

\bibitem[\protect\citeauthoryear{{Fong} et~al.}{{Fong} et~al.}{2013}]{fbc+13}
{Fong}, W., et~al. 2013, \apj, 769, 56

\bibitem[\protect\citeauthoryear{{Fong} et~al.}{{Fong} et~al.}{2011}]{fbc+11}
{Fong}, W., et~al. 2011, \apj, 730, 26

\bibitem[\protect\citeauthoryear{{Fong}, {Berger}, \& {Fox}}{{Fong}
  et~al.}{2010}]{fbf10}
{Fong}, W., {Berger}, E.,  \& {Fox}, D.~B. 2010, \apj, 708, 9

\bibitem[\protect\citeauthoryear{{Fong} et~al.}{{Fong} et~al.}{2012}]{fbm+12}
{Fong}, W., et~al. 2012, \apj, 756, 189

\bibitem[\protect\citeauthoryear{{Forbes} \& {Ponman}}{{Forbes} \&
  {Ponman}}{1999}]{fp99}
{Forbes}, D.~A.,  \& {Ponman}, T.~J. 1999, \mnras, 309, 623

\bibitem[\protect\citeauthoryear{{Fruchter} et~al.}{{Fruchter}
  et~al.}{2006}]{fls+06}
{Fruchter}, A.~S., et~al. 2006, \nat, 441, 463

\bibitem[\protect\citeauthoryear{{Fryer} \& {Kalogera}}{{Fryer} \&
  {Kalogera}}{1997}]{fk97}
{Fryer}, C.,  \& {Kalogera}, V. 1997, \apj, 489, 244

\bibitem[\protect\citeauthoryear{{Fryer}, {Woosley}, \& {Hartmann}}{{Fryer}
  et~al.}{1999}]{fwh99}
{Fryer}, C.~L., {Woosley}, S.~E.,  \& {Hartmann}, D.~H. 1999, \apj, 526, 152

\bibitem[\protect\citeauthoryear{{Galbany} et~al.}{{Galbany}
  et~al.}{2012}]{gmo+12}
{Galbany}, L., et~al. 2012, \apj, 755, 125

\bibitem[\protect\citeauthoryear{{Goad} et~al.}{{Goad} et~al.}{2007}]{gtb+07}
{Goad}, M.~R., et~al. 2007, \aap, 476, 1401

\bibitem[\protect\citeauthoryear{{Graham} et~al.}{{Graham}
  et~al.}{2009}]{gfl+09}
{Graham}, J.~F., et~al. 2009, \apj, 698, 1620

\bibitem[\protect\citeauthoryear{{Hakobyan} et~al.}{{Hakobyan}
  et~al.}{2008}]{hpm+08}
{Hakobyan}, A.~A., {Petrosian}, A.~R., {McLean}, B., {Kunth}, D., {Allen},
  R.~J., {Turatto}, M.,  \& {Barbon}, R. 2008, \aap, 488, 523

\bibitem[\protect\citeauthoryear{{Hill} et~al.}{{Hill} et~al.}{2010}]{hdc+10}
{Hill}, D.~T., {Driver}, S.~P., {Cameron}, E., {Cross}, N., {Liske}, J.,  \&
  {Robotham}, A. 2010, \mnras, 404, 1215

\bibitem[\protect\citeauthoryear{{Holland}, {de Pasquale}, \&
  {Markwardt}}{{Holland} et~al.}{2007}]{gcn7145}
{Holland}, S.~T., {de Pasquale}, M.,  \& {Markwardt}, C.~B. 2007, GRB
  Coordinates Network, 7145, 1

\bibitem[\protect\citeauthoryear{{Kelly} \& {Kirshner}}{{Kelly} \&
  {Kirshner}}{2012}]{kk12}
{Kelly}, P.~L.,  \& {Kirshner}, R.~P. 2012, \apj, 759, 107

\bibitem[\protect\citeauthoryear{{Kelly}, {Kirshner}, \& {Pahre}}{{Kelly}
  et~al.}{2008}]{kkp08}
{Kelly}, P.~L., {Kirshner}, R.~P.,  \& {Pahre}, M. 2008, \apj, 687, 1201

\bibitem[\protect\citeauthoryear{{Kocevski} et~al.}{{Kocevski}
  et~al.}{2010}]{ktr+10}
{Kocevski}, D., et~al. 2010, \mnras, 404, 963

\bibitem[\protect\citeauthoryear{{Kouveliotou} et~al.}{{Kouveliotou}
  et~al.}{1993}]{kmf+93}
{Kouveliotou}, C., {Meegan}, C.~A., {Fishman}, G.~J., {Bhat}, N.~P., {Briggs},
  M.~S., {Koshut}, T.~M., {Paciesas}, W.~S.,  \& {Pendleton}, G.~N. 1993,
  \apjl, 413, L101

\bibitem[\protect\citeauthoryear{{Landsman}, {Marshall}, \&
  {Racusin}}{{Landsman} et~al.}{2007}]{gcn6689}
{Landsman}, W., {Marshall}, F.~E.,  \& {Racusin}, J. 2007, GRB Coordinates
  Network, 6689, 1

\bibitem[\protect\citeauthoryear{{Leibler} \& {Berger}}{{Leibler} \&
  {Berger}}{2010}]{lb10}
{Leibler}, C.~N.,  \& {Berger}, E. 2010, \apj, 725, 1202

\bibitem[\protect\citeauthoryear{{Levan} et~al.}{{Levan}
  et~al.}{2009}]{gcn10154}
{Levan}, A.~J., {Tanvir}, N.~R., {Hjorth}, J., {Malesani}, D., {de Ugarte
  Postigo}, A.,  \& {D'Avanzo}, P. 2009, GRB Coordinates Network, 10154, 1

\bibitem[\protect\citeauthoryear{{Levan} et~al.}{{Levan} et~al.}{2006}]{lwc+06}
{Levan}, A.~J., {Wynn}, G.~A., {Chapman}, R., {Davies}, M.~B., {King}, A.~R.,
  {Priddey}, R.~S.,  \& {Tanvir}, N.~R. 2006, \mnras, 368, L1

\bibitem[\protect\citeauthoryear{{Levesque} et~al.}{{Levesque}
  et~al.}{2010}]{lbb+10}
{Levesque}, E.~M., et~al. 2010, \mnras, 401, 963

\bibitem[\protect\citeauthoryear{{Li} et~al.}{{Li} et~al.}{2011}]{lcl+11}
{Li}, W., {Chornock}, R., {Leaman}, J., {Filippenko}, A.~V., {Poznanski}, D.,
  {Wang}, X., {Ganeshalingam}, M.,  \& {Mannucci}, F. 2011, \mnras, 412, 1473

\bibitem[\protect\citeauthoryear{{Mannucci} et~al.}{{Mannucci}
  et~al.}{2005}]{mdp+05}
{Mannucci}, F., {Della Valle}, M., {Panagia}, N., {Cappellaro}, E., {Cresci},
  G., {Maiolino}, R., {Petrosian}, A.,  \& {Turatto}, M. 2005, \aap, 433, 807

\bibitem[\protect\citeauthoryear{{Mao} et~al.}{{Mao} et~al.}{2008}]{gcnr138}
{Mao}, J., {Guidorzi}, C., {Ukwatta}, T., {Brown}, P.~J., {Barthelmy}, S.~D.,
  {Burrows}, D.~N., {Roming}, P.,  \& {Gehrels}, N. 2008, GCN Report, 138, 1

\bibitem[\protect\citeauthoryear{{Marchesini} et~al.}{{Marchesini}
  et~al.}{2012}]{msb+12}
{Marchesini}, D., {Stefanon}, M., {Brammer}, G.~B.,  \& {Whitaker}, K.~E. 2012,
  \apj, 748, 126

\bibitem[\protect\citeauthoryear{{Marchesini} et~al.}{{Marchesini}
  et~al.}{2007}]{mvq+07}
{Marchesini}, D., et~al. 2007, \apj, 656, 42

\bibitem[\protect\citeauthoryear{{Margutti} et~al.}{{Margutti}
  et~al.}{2012}]{mbf+12}
{Margutti}, R., et~al. 2012, \apj, 756, 63

\bibitem[\protect\citeauthoryear{{McBreen} et~al.}{{McBreen}
  et~al.}{2010}]{mkr+10}
{McBreen}, S., et~al. 2010, \aap, 516, A71

\bibitem[\protect\citeauthoryear{{Metzger}, {Quataert}, \&
  {Thompson}}{{Metzger} et~al.}{2008}]{mqt08}
{Metzger}, B.~D., {Quataert}, E.,  \& {Thompson}, T.~A. 2008, \mnras, 385, 1455

\bibitem[\protect\citeauthoryear{{Narayan}, {Paczynski}, \& {Piran}}{{Narayan}
  et~al.}{1992}]{npp92}
{Narayan}, R., {Paczynski}, B.,  \& {Piran}, T. 1992, \apjl, 395, L83

\bibitem[\protect\citeauthoryear{{Nicuesa Guelbenzu} et~al.}{{Nicuesa
  Guelbenzu} et~al.}{2012}]{nkk+12}
{Nicuesa Guelbenzu}, A., et~al. 2012, \aap, 538, L7

\bibitem[\protect\citeauthoryear{{Oemler} \& {Tinsley}}{{Oemler} \&
  {Tinsley}}{1979}]{ot79}
{Oemler}, A., Jr.,  \& {Tinsley}, B.~M. 1979, \aj, 84, 985

\bibitem[\protect\citeauthoryear{{Perley} et~al.}{{Perley}
  et~al.}{2008}]{pbm+08}
{Perley}, D.~A., {Bloom}, J.~S., {Modjaz}, M., {Miller}, A.~A., {Shiode}, J.,
  {Brewer}, J., {Starr}, D.,  \& {Kennedy}, R. 2008, GRB Coordinates Network,
  7889, 1

\bibitem[\protect\citeauthoryear{{Perley} et~al.}{{Perley}
  et~al.}{2009}]{pmg+09}
{Perley}, D.~A., et~al. 2009, \apj, 696, 1871

\bibitem[\protect\citeauthoryear{{Perley} et~al.}{{Perley}
  et~al.}{2012}]{pmm+12}
{Perley}, D.~A., {Modjaz}, M., {Morgan}, A.~N., {Cenko}, S.~B., {Bloom}, J.~S.,
  {Butler}, N.~R., {Filippenko}, A.~V.,  \& {Miller}, A.~A. 2012, \apj, 758,
  122

\bibitem[\protect\citeauthoryear{{Perley} et~al.}{{Perley}
  et~al.}{2007}]{ptc+07}
{Perley}, D.~A., {Thoene}, C.~C., {Cooke}, J., {Bloom}, J.~S.,  \& {Barton}, E.
  2007, GRB Coordinates Network, 6739, 1

\bibitem[\protect\citeauthoryear{{Piranomonte} et~al.}{{Piranomonte}
  et~al.}{2008}]{pdc+08}
{Piranomonte}, S., et~al. 2008, \aap, 491, 183

\bibitem[\protect\citeauthoryear{{Poli} et~al.}{{Poli} et~al.}{2003}]{pgf+03}
{Poli}, F., et~al. 2003, \apjl, 593, L1

\bibitem[\protect\citeauthoryear{{Prieto}, {Stanek}, \& {Beacom}}{{Prieto}
  et~al.}{2008}]{psb08}
{Prieto}, J.~L., {Stanek}, K.~Z.,  \& {Beacom}, J.~F. 2008, \apj, 673, 999

\bibitem[\protect\citeauthoryear{{Qin} et~al.}{{Qin} et~al.}{1998}]{qwc+98}
{Qin}, B., {Wu}, X.-P., {Chu}, M.-C., {Fang}, L.-Z.,  \& {Hu}, J.-Y. 1998,
  \apjl, 494, L57

\bibitem[\protect\citeauthoryear{{Racusin}, {Barbier}, \& {Landsman}}{{Racusin}
  et~al.}{2007}]{gcnr70}
{Racusin}, J., {Barbier}, L.,  \& {Landsman}, W. 2007, GCN Report, 70, 1

\bibitem[\protect\citeauthoryear{{Ramos} et~al.}{{Ramos} et~al.}{2011}]{rpb+11}
{Ramos}, B.~H.~F., et~al. 2011, \aj, 142, 41

\bibitem[\protect\citeauthoryear{{Rosswog}, {Ramirez-Ruiz}, \&
  {Davies}}{{Rosswog} et~al.}{2003}]{rrd03}
{Rosswog}, S., {Ramirez-Ruiz}, E.,  \& {Davies}, M.~B. 2003, \mnras, 345, 1077

\bibitem[\protect\citeauthoryear{{Rowlinson} et~al.}{{Rowlinson}
  et~al.}{2010a}]{rot+10}
{Rowlinson}, A., et~al. 2010a, \mnras, 409, 531

\bibitem[\protect\citeauthoryear{{Rowlinson} et~al.}{{Rowlinson}
  et~al.}{2010b}]{rwl+10}
{Rowlinson}, A., et~al. 2010b, \mnras, 408, 383

\bibitem[\protect\citeauthoryear{{Sakamoto} et~al.}{{Sakamoto}
  et~al.}{2013}]{sta+12}
{Sakamoto}, T., et~al. 2013, \apj, 766, 41

\bibitem[\protect\citeauthoryear{{Saracco} et~al.}{{Saracco}
  et~al.}{2006}]{sfc+06}
{Saracco}, P., et~al. 2006, \mnras, 367, 349

\bibitem[\protect\citeauthoryear{{Schlafly} \& {Finkbeiner}}{{Schlafly} \&
  {Finkbeiner}}{2011}]{sf11}
{Schlafly}, E.~F.,  \& {Finkbeiner}, D.~P. 2011, \apj, 737, 103

\bibitem[\protect\citeauthoryear{{Soderberg} et~al.}{{Soderberg}
  et~al.}{2006}]{sbk+06}
{Soderberg}, A.~M., et~al. 2006, \apj, 650, 261

\bibitem[\protect\citeauthoryear{{Stefanon} \& {Marchesini}}{{Stefanon} \&
  {Marchesini}}{2013}]{sm13}
{Stefanon}, M.,  \& {Marchesini}, D. 2013, \mnras, 429, 881

\bibitem[\protect\citeauthoryear{{Stratta} et~al.}{{Stratta}
  et~al.}{2007}]{sdp+07}
{Stratta}, G., et~al. 2007, \aap, 474, 827

\bibitem[\protect\citeauthoryear{{Svensson} et~al.}{{Svensson}
  et~al.}{2010}]{slt+10}
{Svensson}, K.~M., {Levan}, A.~J., {Tanvir}, N.~R., {Fruchter}, A.~S.,  \&
  {Strolger}, L.-G. 2010, \mnras, 405, 57

\bibitem[\protect\citeauthoryear{{Tanvir} et~al.}{{Tanvir}
  et~al.}{2013}]{tlf+13}
{Tanvir}, N.~R., {Levan}, A.~J., {Fruchter}, A.~S., {Hjorth}, J., {Wiersema},
  K., {Tunnicliffe}, R.,  \& {de Ugarte Postigo}, A. 2013, ArXiv e-prints

\bibitem[\protect\citeauthoryear{{Troja} et~al.}{{Troja} et~al.}{2008}]{tko+08}
{Troja}, E., {King}, A.~R., {O'Brien}, P.~T., {Lyons}, N.,  \& {Cusumano}, G.
  2008, \mnras, 385, L10

\bibitem[\protect\citeauthoryear{{van den Bergh}, {Li}, \& {Filippenko}}{{van
  den Bergh} et~al.}{2005}]{vlf05}
{van den Bergh}, S., {Li}, W.,  \& {Filippenko}, A.~V. 2005, \pasp, 117, 773

\bibitem[\protect\citeauthoryear{{Wainwright}, {Berger}, \&
  {Penprase}}{{Wainwright} et~al.}{2007}]{wbp07}
{Wainwright}, C., {Berger}, E.,  \& {Penprase}, B.~E. 2007, \apj, 657, 367

\bibitem[\protect\citeauthoryear{{Wang}, {Lai}, \& {Han}}{{Wang}
  et~al.}{2006}]{wlh+06}
{Wang}, C., {Lai}, D.,  \& {Han}, J.~L. 2006, \apj, 639, 1007

\bibitem[\protect\citeauthoryear{{Wang} et~al.}{{Wang} et~al.}{2013}]{wwf+13}
{Wang}, X., {Wang}, L., {Filippenko}, A.~V., {Zhang}, T.,  \& {Zhao}, X. 2013,
  ArXiv e-prints

\bibitem[\protect\citeauthoryear{{Wong}, {Willems}, \& {Kalogera}}{{Wong}
  et~al.}{2010}]{wwk10}
{Wong}, T.-W., {Willems}, B.,  \& {Kalogera}, V. 2010, \apj, 721, 1689

\bibitem[\protect\citeauthoryear{{Xue} et~al.}{{Xue} et~al.}{2008}]{xrz+08}
{Xue}, X.~X., et~al. 2008, \apj, 684, 1143

\bibitem[\protect\citeauthoryear{{Zemp}, {Ramirez-Ruiz}, \& {Diemand}}{{Zemp}
  et~al.}{2009}]{zrd09}
{Zemp}, M., {Ramirez-Ruiz}, E.,  \& {Diemand}, J. 2009, \apjl, 705, L186

\end{thebibliography}

\end{document}